# Virtual Molecular Dynamics

Yuriy V. Sereda[1], Andrew Abi Mansour,[2] and Peter J. Ortoleva[1]*

[a]Department of Chemistry

Indiana University

Bloomington, IN 47405

*Contact: *ortoleva@indiana.edu*

[2]Center for Materials Science and Engineering,

Merck and Co., Inc., West Point, PA 19486

**Abstract**

Molecular dynamics is based on solving Newton's equations for many-particle systems that evolve along complex, highly fluctuating trajectories. The orbital instability and short-time complexity of Newtonian orbits is in sharp contrast to the more coherent behavior of collective modes such as density profiles. The notion of virtual molecular dynamics is introduced here based on temporal coarse-graining via Padé approximants and the Ito formula for stochastic processes. It is demonstrated that this framework leads to significant efficiency over traditional molecular dynamics and avoids the need to introduce coarse-grained variables and phenomenological equations for their evolution. In this framework, an all-atom trajectory is represented by a Markov chain of virtual atomic states at a discrete sequence of timesteps, transitions between which are determined by an integration of conventional molecular dynamics with Padé approximants and a microstate energy annealing methodology. The latter is achieved by a conventional and an MD NVE energy minimization schemes. This multiscale framework is demonstrated for a pertussis toxin subunit undergoing a structural transition, a T=1 capsid-like structure of HPV16 L1 protein, and two coalescing argon droplets.

**I. Introduction**

Molecular dynamics (MD) is the evolution of many-atom systems over time via Newton's equations and an interatomic force field. The deterministic nature of Newtonian evolution of a many-atom system is masked by its inherent complexity. The Newtonian trajectories are unstable in the dynamical sense, and therefore there is frequent switching between the trajectories corresponding to close initial conditions.[1] The objective of this study is to introduce the notion that MD can be treated as a multiscale stochastic process with transition probability between discrete states determined by the rapidly fluctuating Newtonian many-atom dynamics. Atomic trajectories generated by this temporal coarse-grained (CG) stochastic process are virtual, i.e., not Newtonian. However, they do capture essential details of the overall motion, i.e., they are consistent with an ensemble of Newtonian trajectories. Hence the approach considered here is termed "virtual MD". It allows advancing the atom-resolved state over large time intervals while maintaining atomic resolution and avoids the need to introduce CG variables and conjecturing phenomenological CG dynamical equations and calibrating them. It is demonstrated here that this perspective holds great promise for efficient all-atom MD simulation. The issue of judging the accuracy of the multiscale simulations relative to conventional MD has been addressed by resorting to ensemble MD,[2] and notably by accounting for indistinguishability of particles for simple systems like argon (as suggested by a notion of identical particles from quantum theory). In the present study, for practical purposes five MD simulations is taken to be the ensemble.

Efficiency in simulating the dynamics of many-atom systems is strongly dependent on the timestep size used to guarantee stability and accuracy. A method based on Padé approximants (PA) for increasing time integration step size, $\Delta$, is developed here. To the best of our knowledge, this is the first application of Padé extrapolation to solving Newton's equations of motion for many-particle systems. PA has been used in many fields of mathematics,[3] biology and chemistry,[4] physics and engineering[4c, 5] to extend the range of approximations (e.g., to longer times, distances[4c] and densities,[6] or from linear to nonlinear regime[4a]) and to determine critical points and system behavior in their vicinity.[7] PA is also a way of integrating known limiting behaviors[8] or experimental information[9] into the solution of a problem. They have been used in both deterministic differential equation models[3b-d, 10] and for the description of stochastic processes, e.g., in electronic circuits[11], control theory,[12] cellular networks,[13] and economics (e.g., for modeling and forecasting multivariate time series[14]). PA extrapolation techniques were extensively tested in molecular physics, such as in the construction of a unified analytical representation of molecular potential energy curves for the radial Schrodinger equation of diatomic molecules,[15] in approximating correlation energy functional for atoms and diatomic molecules with self-consistent local density approximation for exchange energy,[16] and in the studies of molecular motions in condensed phases by constructing closed-form approximation to orientational correlations.[17]

To achieve timesteps large relative to conventional MD, a PA scheme should account for the stochastic nature of MD. A special formulation of the PA guided by the Ito formula[18] from the theory of random processes is developed here. Optimal efficiency is achieved by also accounting for the stationarity concept,[19] thereby only requiring a short conventional MD simulation to advance the system by a large timestep. Efficiency of the method rests on choosing the form of the PA so that the CG description it embodies varies slowly in time, and thereby enabling extrapolation over long timesteps. In particular, the all-atom state predicted should not require extensive energy minimization and thermal equilibration, which otherwise would diminish efficiency of the method.

The remainder of this paper is arranged as follows. In Sec. II, the proposed virtual MD algorithm is presented. In Sec. III, the method is applied to bonded and non-bonded molecular systems covering a broad range of sizes. Discussion and conclusions are presented in Sec. IV.

## II. Materials and Methods

It is difficult to predict the atom-resolved details of complex atomic motions over large timesteps $\Delta$. To address this, the notion of virtual atomic trajectories is introduced in a Markovian approach which involves a transition probability that takes the system from a state at $n\Delta$ to one at $(n+1)\Delta$, $(n=1,2,...)$. The transition probability incorporates key aspects of Newtonian mechanics via a set of short MD simulations, each of duration $\delta\,(\ll\Delta)$. However, since $\Delta$ is large, a virtual trajectory does not evolve by Newton's equations directly. The transition of the configuration from that at time $n\Delta$ to one at time $(n+1)\Delta$ is guided by historical information prior to $n\Delta$, as well as that from a short MD run starting at $n\Delta$. Two sets of all-atom configurations are taken from two latest MD simulations (the previous MD interval starting at time $(n-1)\Delta$, and the latest starting at $n\Delta$) to construct a PA for $t=(n+1)\Delta$. The resulting atomic displacements are constrained to avoid unphysical configurations with overlapping atoms. A sufficient number of all-atom configurations has to be taken from each such MD time interval to ensure statistical significance of the resulting Markovian trajectory. This allows to avoid the problem of exponential divergence of MD trajectories.[1]

Temporal evolution in the proposed virtual MD method is achieved via a sequence of three-stage timesteps (Fig. 1). In the first stage an ensemble of atomic positions is generated by a short conventional MD simulation of duration $\delta$. Next a time coarse-graining of this MD trajectory is achieved by constructing a PA for each of the $3N$ atomic coordinates to extrapolate positions of atoms and advance the system from time $n\Delta$ to $(n+1)\Delta$. PAs are constructed by finding their coefficients via a least square approach using two most recent MD ensembles. To preserve the stochastic aspects of the trajectory in the PA, the latter is expressed so as to account for the Ito formula, as earlier.[20] In the last optional stage of advancing the system over a virtual MD timestep, the PA-extrapolated all-atom configuration (termed a "PA configuration" below for brevity) is annealed by relieving unphysical deformations of bond lengths and angles, and thermally equilibrated when necessary. In the present applications of virtual MD, annealing of the microstate is done by conventional energy minimization scheme using a combination of steepest descent and conjugate gradient methods. The resulting all-atom configuration is termed an "EM configuration". When the minimization converges to a microstate with high potential energy, so that its subsequent thermal equilibration or MD run is numerically unstable, additional energy minimization is performed using MD NVE (Sect. III). The resulting structure is termed an "NVE configuration". If NVE simulation also cannot be started due to high energy, then Gromacs minimization parameters are adjusted to improve the energy minimization. The NVE configuration is optionally thermally equilibrated to complete the CG virtual MD simulation step. The equilibrated all-atom configuration, termed an "EQ configuration", is used as input for the next virtual MD step. Real applications require multiple CG virtual MD steps to be performed.

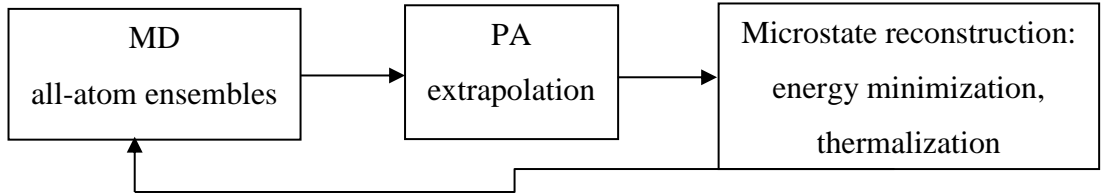

Fig. 1. Workflow of virtual MD. Each CG simulation step in virtual MD consists of three stages: generation of all-atom MD ensemble to be used in the PA extrapolation stage, followed by microstate reconstruction using energy minimization and thermal equilibration when necessary. The resulting virtual atomic configuration must have energy low enough to enable subsequent virtual MD step.

The classic PA has the form $A(\tau)/B(\tau)$ with $A$ and $B$ being polynomials.[3a] For a many-atom system, atomic motions can be described as a combination of coherent and stochastic components. Here the forms of $A$ and $B$ are conjectured using notions such as the Ito formula and the boundedness of PA. According to the Ito formula, coherent effects grow as $\tau$ and the stochastic ones as $\tau^{1/2}$. Consider an interval of time from $t-2\Delta$ to $t$ in a discrete time-advancement scheme for an arbitrary time $t$. Let $\tau$ be a time relative to the beginning of this interval, i.e., $\tau = 0$ at $t-2\Delta$. In the present scheme, a PA is constructed as

$$P(\tau) = \frac{a_0 + a_{1/2}\tau^{1/2} + a_1\tau}{1+b^2\tau}, \quad 0 < \tau < 2\Delta. \quad (1)$$

The $a_{1/2}$-term tracks fluctuational aspect of atomic motions. While $a_{1/2}$ fluctuates around zero (see Fig. 6 in Ref.[20]), the coefficients $a_0$, $a_1$, and $b_1$ show coherent behavior.

In the current implementation, the sets of four coefficients $a_0$, $a_{1/2}$, $a_1$, and $b$ are to be determined for each of the $3N$ atomic coordinates independently, based on the individual history for each coordinate. This enables maximum possible parallelization of the second PA extrapolation stage of virtual MD. In addition, PA extrapolation (2) is O($N$), unlike MD which is of higher order O($N\log N$).

Non-singular PA with an exponential function in denominator (as earlier[20]) was replaced by the polynomial function to facilitate analytical solving of constraints for the coefficients of the PA (1). The $\tau^{1/2}$ term in numerator of Eq. (1) comes from the Ito formula.[18]

After extensive and unsuccessful experimentation with minimum necessary number of four all-atom MD configurations in one ensemble of states at time interval $n\Delta, n\Delta + \delta$, and with different polynomial and exponential functions in the numerator and denominator, it was concluded that a larger number of all-atom configurations in each MD ensemble is necessary. Eventually, 11 configurations were chosen to be used in each of the two MD ensembles, despite the increased compute time and especially algebraic complexity of the constraints for the four PA coefficients. Interval between historical MD configurations was chosen to be 1 ps in two applications to bonded protein systems, and 0.5 ps for non-bonded argon system (see Table I). This is a typical interval between frames recorded in MD simulations of such systems. Three values of the PA step, $\Delta = 5, 10, 20$ ps, were tested. In order to estimate an upper limit of $\Delta$, and thereby a maximum possible speedup over MD, a characteristic time of system evolution was estimated for one of the application systems (namely, pertussis toxin) using the ratio of first to second derivatives of the PA (1) at the extrapolation time $\tau = 2\Delta$, averaged over all atoms. It was shown that the maximum PA step $\Delta = 20$ ps is well below the characteristic time.

Use of information from previous PA steps, i.e., choosing MD ensembles (at $\tau = 0, 1, \ldots \delta$ and $\tau = \Delta, \Delta+1, \ldots \Delta+\delta$, with $\delta = 10$ ps for proteins and $\delta = 5$ ps for argon) allows one to take larger timestep $\Delta$ and consequently achieve greater computational efficiency. This multistep method is nonlinear, in the sense that the extrapolation is not a linear combination of the coordinates at previous steps. It is an explicit method where a next step can be computed directly from the previous information, and no unknown information about the next step is involved. Although an implicit predictor-corrector method of PA extrapolation was shown to be effective,[20] it was applied to few CG variables therein and, we believe, it would take too much time to compute for the 3*N* atomic coordinates to provide speedup over conventional MD.

Construction of the PA is achieved by the least squares method, where one minimizes the sum of the squares of offsets of the PA, $P(\tau_i)$, from the historical values, $f_i, i = 1, \ldots 22$, of a given atomic coordinate taken from the MD ensembles. In this study, PA coefficients were found by minimizing the sum of squared deviations in a semi-analytical way with Maple$^{TM}$.[21] Namely, three of the PA coefficients were expressed via a remaining one for a given set of numerical values of $f_i$. The remaining last equation is solved numerically.

The focus of the present study is to show feasibility of the proposed PA-accelerated virtual MD approach, and to elaborate preliminary workflow for its stages. With this goal in mind, only one virtual MD step was launched starting from a given set of 22 historical all-atom MD configurations. However, in a real application of the virtual MD, the minimized and equilibrated PA configuration would be used to initiate next virtual MD step by launching short MD simulation from it. Numerous sets of initial MD configurations were considered for each of the three demonstration systems. In each of these sets, 22 all-atom states were taken at equal time intervals within two MD intervals (each of duration $\delta$) throughout a long conventional MD trajectory for a given system. To estimate the computational speedup, the energy minimization and thermal equilibration stages were shortened, while still achieving microstate which could serve as an initial condition for subsequent MD run.

Comparison of the traditional MD trajectory projected to *xy* plane for one of the argon droplet atoms with its PA extrapolations to $2\Delta, 3\Delta,...$ (without microstate annealing) in Fig. 2 shows that the PA-extrapolated virtual trajectory follows the trend of conventional MD simulation.

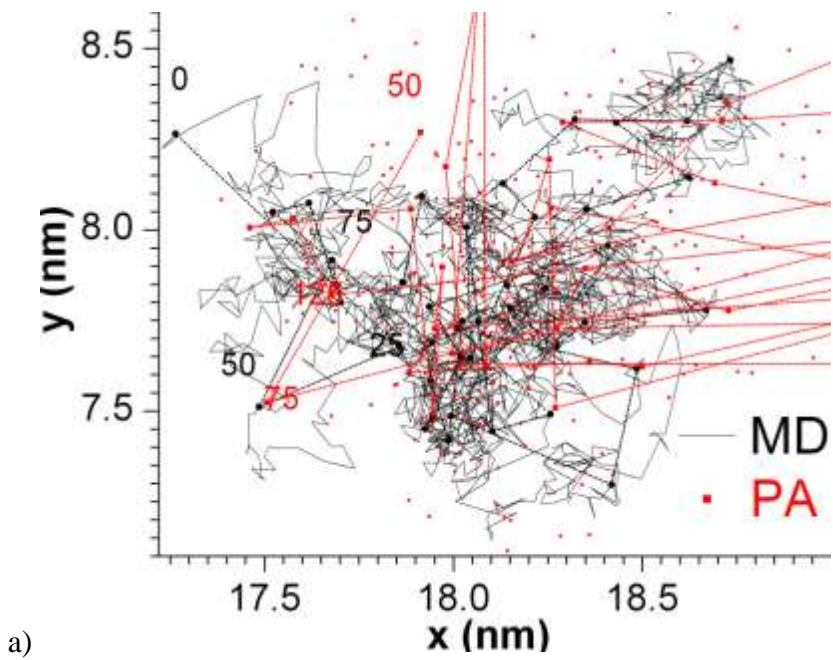

a)

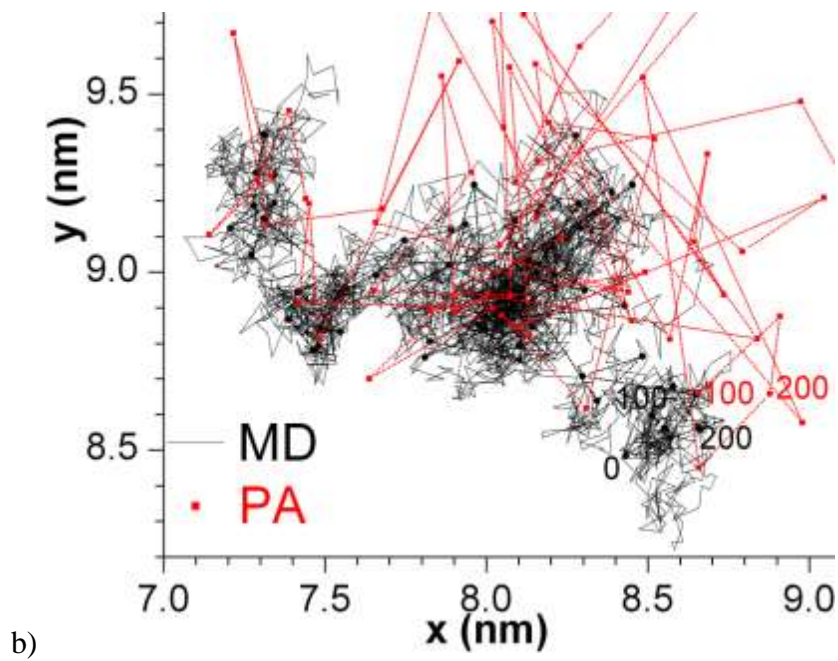

b)

Fig. 2. Direct comparison of the position of a selected atom in the conventional MD NVT simulation with its one-step extrapolation generated by the PA for a) argon droplets coalescence, and b) 1PRT protein in solvent. (*Black*) Newtonian trajectory generated in the conventional MD simulation, which is a discretized solution of Newton's equations, with small enough timestep (0.5 fs) so that deviations from the continuous trajectory are minimal. Points at every $\Delta = 25$ ps are connected by dotted line for easier comparison to the PA extrapolations. (*Red*) The virtual MD trajectory is, by construction, only known at a set of discrete time points $2\Delta, 3\Delta, ...$, (large squares connected by *red lines*) where timestep $\Delta$ is much greater than that used in conventional MD (0.5 fs). (*Small red squares*) PA configurations constructed every 2.5 ps. *Labels* show time in ps. To generate atomic coordinates at $t = (n+1)\Delta$, atomic configurations from two MD simulations at time intervals $(n-1)\Delta, (n-1)\Delta + \delta$ and $n\Delta, n\Delta + \delta$ are used, with $\Delta = 25$ps and $\delta = 5$ps. Lines connecting the discrete states are purely for visualization to suggest there is a temporal sequence.

Performance of the algorithm is assessed using all-atom conventional MD trajectories of (a) structural fluctuations in pertussis toxin (PDB code 1PRT)[22], (b) shrinking of HPV16 T=1 capsid-like structure in vacuum (the latter two are bonded protein systems), and (c) coalescence of two argon nanodroplets (non-bonded system,[23] Fig. 3). All conventional MD simulations (including MD run used for taking historical information on atomic positions, as well as ensemble generation, minimizations, and equilibration in virtual MD) were performed using Gromacs[24] with CHARMM27 force field.[25] MD trajectory for HPV16 T=1 capsid-like structure and argon droplets was prepared using NVT ensemble with velocity-rescale modified Berendsen temperature coupling.[26] MD trajectory for 1PRT protein was prepared using NPT ensemble with Berendsen pressure coupling.[27] Other MD parameters used in present simulations are shown in Table I. A description of each application system is given below.

**Table I**. Conventional MD parameters for three application systems.

| System | $N_{atoms}$ | Solvent Model | dt [ps] | Van der Waals cutoff [nm] | Coulomb cutoff [nm] | Temperature [K] | NVT box [nm$^3$] |
|---|---|---|---|---|---|---|---|
| 1PRT | 18,637 | Explicit (TIP3P) | 1 | 1.4 | 1.4 | 300 | 16 x 16 x 24 |
| HPV16 | 473,640 | vacuum | 1 | 1.4 | 1.4 | 300 | $37 \times 37 \times 37$ |
| Argon | 87,820 | vacuum | 0.5 | 1.2 | 1.2 | 70 | $36 \times 20 \times 20$ |

Pertussis toxin (PDB code 1PRT), a protein-based exotoxin,[22] was used here to illustrate application of virtual MD to bonded systems and, in particular, average-size proteins. Na and Cl counter-ions of concentration 0.15 M were added for charge neutrality. The ionized and solvated system consisted of 603,775 atoms. An equilibration run with position restraints imposed on the protein was performed for 100 ps; after thermal equilibrium was established, the system was simulated without any restraints for 3.2 ns during which the protein underwent a conformational change involving local compression-extension, as reflected by its mass density.[20]

The second simulated bonded system is a T=1 icosahedral assembly of twelve L1 pentamers of the human papillomavirus (HPV16, PDB code 1DZL).[28] This virus-like particle was chosen to test capability of virtual MD to accelerate MD of biomolecular and other large-size nanosystems. This system supports a stable T=1 structure, particularly upon truncation of L1 protein residues.[28] Initially, the system was solvated, ionized with 0.15 M NaCl ions, charge neutralized, and then thermally equilibrated. However, due to its large size (see Table I) and a large number of simulations with varied sets of MD parameters needed in this exploratory study, the simulations were performed in vacuum. HPV16 T=1 capsid-like structure was shrinking throughout 16 ns of its conventional MD simulation.

Next the virtual MD is tested on a non-bonded system of two coalescing argon nanodroplets.[23, 29] Nanodroplets are nucleated in association with changes in thermochemical conditions such as those accompanying the natural or engineered generation of hydrocarbon liquid phases. After early droplet nucleation and growth, Ostwald ripening (mediated by diffusional exchange of monomers) or coalescence leads to fewer and larger droplet sizes. Fig. 3 illustrates the coalescence of two argon nanodroplets, initially (before thermal equilibration) each of radius 7.5 nm and separated by a minimum distance of 1 nm. The initial density of each droplet is 1.65 g/ml. Preparation of the droplets is described in the Supplementary Information. A production MD NVT run is performed for 1 ns. At time t ≈ 10 ps, the two droplets begin to form a bridge induced by the overlapping Van der Waals attractive tails. Then, the bridge thickens to decrease the high curvature of the connection region until the system reaches thermodynamic equilibrium when the nanodroplets form a united and larger droplet. Such phenomena are not readily simulated with continuum models unless a rheological equation is postulated that accounts for the dramatic change in spatial scales.[30] For example, when the two droplets first touch, there is a region of infinite curvature at which continuum models break down (although such effects are crudely accounted for in a Landau-type model[31]). Such high-curvature effects are accounted for in the present PA scheme.

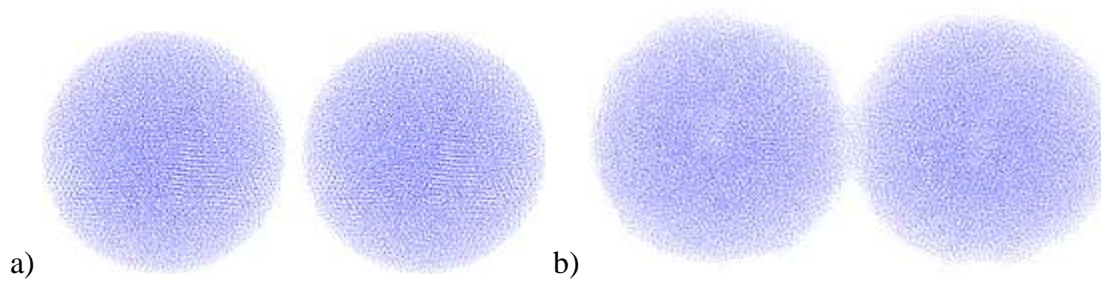

a)                          b)

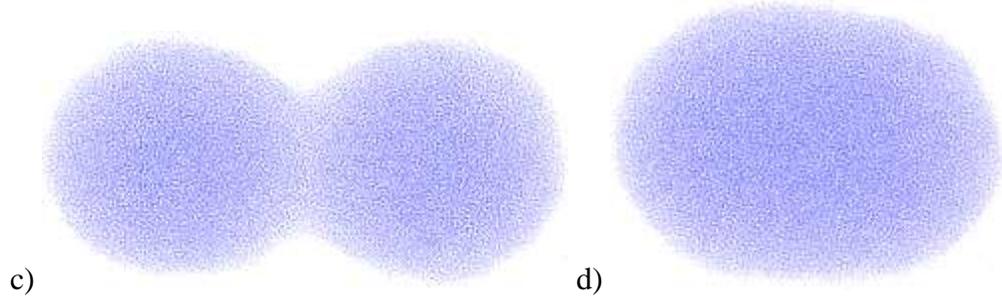

c)            d)

Fig. 3. Coalescence of two argon nanodroplets in conventional MD NVT simulation in vacuum. a) Two droplets initially 1 nm apart from each other, b) bridge formation by 20 ps, c) dumbbell-shaped structure at 100 ps, and d) merged droplet at 500 ps. A small portion of the atoms disperses, while most of the atoms form a droplet that is held together by short-range Van der Waals forces. By 1 ns, the merged droplet becomes almost spherical.

## III. Results

An assessment of the virtual MD algorithm (Sect. II) is made using three application systems of different size and bonding character. This assessment was performed at different stages of system evolution by making one virtual MD step starting from a set of discrete and equally-spaced time points in the conventional MD simulation. To enhance the statistics underlying our predictions for atomic positions in a given system, in the first MD stage of virtual MD, a historical information on atomic coordinates is collected at two conventional MD simulation intervals starting at time points $t-2\Delta$ and in the considered point $t-\Delta$, each of duration $\delta < \Delta$, from each of which 11 all-atom configurations are chosen at equal time intervals $\delta/10$. In subsequent stages of a virtual MD step, advancement of a given 22-point ensemble of microstates is performed from time $t-\Delta$ to time $t$ (Sect. II).

       The detailed protocol for each stage of the virtual MD step is as follows. Transformation of the all-atom configuration through PA extrapolation, energy minimization, and thermal equilibration stages was used to advance the system to time $t$ (Fig. 1). PA, EM and EQ configurations were compared to the configuration predicted by conventional MD at the extrapolated time $t$. Potential energy, radius of gyration, and RMSD of the virtual MD configurations relative to that from conventional MD show the dependence of the accuracy of virtual MD on the ratio $\Delta/\delta$ and the amount of minimization and equilibration. In addition, EQ

configuration was compared to entire conventional MD trajectory using its RMSD relative to a set of conventional MD configurations at different stages of system evolution. This analysis shows that thermal equilibration in virtual MD adds unphysical evolution to the EM configuration in the sense that the EQ configuration starts resembling conventional MD configurations from times later than $t$ more than that from earlier times. Therefore, for the sake of accuracy and speedup it is imperative to do not have extra equilibration beyond an amount sufficient to arrive at configuration that can serve as input to conventional MD, thus enabling subsequent virtual MD step. Comparison of RMSD is the most rigid test used since the Newtonian orbits are not stable and tend to diverge over long times (see Sect. III).

To demonstrate a speedup of virtual MD versus conventional MD, and to find the ways to increase it, three values of the $\Delta/\delta$ ratio were studied, $\Delta/\delta = 5, 10, 20$ for 1PRT protein. One value, $\Delta/\delta = 5$, was studied for HPV16 T=1 capsid-like structure and argon droplets to demonstrate feasibility of virtual MD for larger bonded and non-bonded systems. Each value of $\Delta/\delta$ required individual adjustment of the MD parameters and durations of minimization and equilibration stages in virtual MD to enable these stages, and increase the speedup. The following maximum speedups were achieved: 17.3 for 1PRT in aqueous solvent, 1.56 for HPV16 T=1 capsid-like structure in vacuum, and 11.93 for argon droplets. A larger speedup could be achieved for HPV16 system if MD parameters were further optimized.

The above mentioned detailed analysis showed ways to improve quality and performance of virtual MD. For example, variations of duration for microstate reconstruction stages of virtual MD allowed to increase speedup, as well as to reveal directions of possible additional speedup. Thus, for 1PRT it was shown that $\Delta/\delta$ could be increased further, since the maximum value of the CG timestep, $\Delta = 20 ps$, is much shorter than the characteristic time of evolution for this system. The characteristic time was found by virtual MD to be longer than 200 ps; such large integration timesteps cannot be achieved in conventional MD. This suggests that additional order of magnitude in efficiency of virtual MD could be achieved by using larger $\Delta$ after proper adjustment of MD parameters. Estimate of the characteristic evolution time has to be updated in the course of a multistep virtual MD simulation since often the qualitative character of evolution changes. For instance, in the considered example of 1PRT structural transition, the characteristic time grows linearly, as found by an analysis of PA configurations generated starting at different points in conventional MD. As a result, testing each stage of virtual MD allowed to obtain sets of MD parameters suitable for realistic, multistep virtual MD simulations for the three tested systems. Further details for each tested system are provided below.

**Pertussis Toxin Protein**

Potential energy (per atom) for the conventional MD configurations of pertussis toxin protein[22] (Fig. S1) is considerably lower than that of the non-bonded argon droplet system (Fig. S10) due to the negative contribution of electrostatic interactions in the bonded system. The same comparison holds for the PA configurations (compare Figs. S2 and S11) due to a volatile character of argon atom motions. The bonded character of 1PRT protein leads to more regular atomic positions in conventional MD stage compared to that for the non-bonded argon system and to fewer steric clashes in PA conformations.

Fig. 4 shows that as $\Delta/\delta$ grows, the difference between PA and conventional MD configurations grows, as expected.

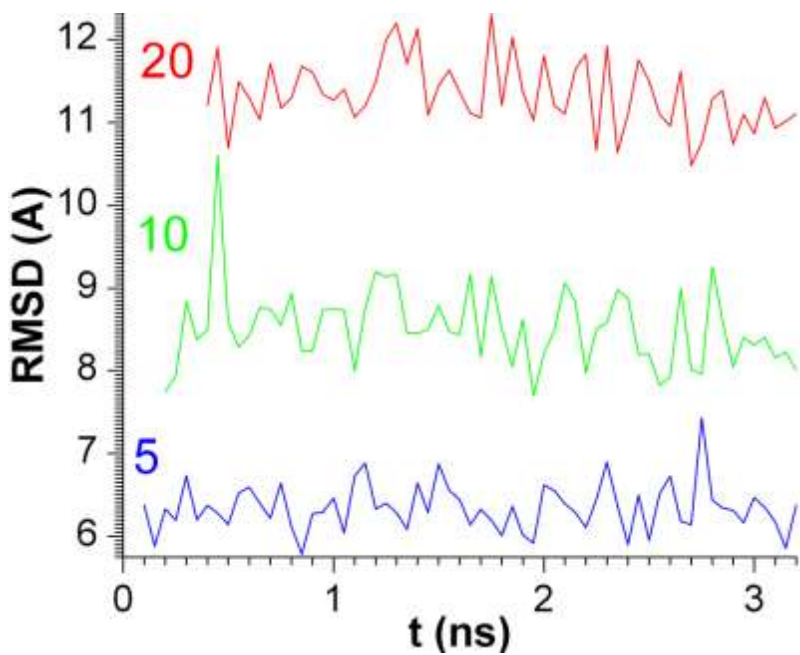

Fig. 4. RMSD between MD and PA configurations at the same extrapolated time. $\Delta/\delta$ values are discriminated by *colors*, as *labeled* on the plot.

Fig. S3 shows that $U(t)$ for the bonded protein system pertussis toxin after energy minimization of the PA configurations is much higher than in MD, in contrast to the non-bonded argon nanodroplet system. A problem arises that some of the PA configurations cannot be effectively energy minimized with the conventionally used set of Gromacs parameters (as in Table I). As in the case of the non-bonded system, at increasing $\Delta/\delta$ value in the PA stage, the $U$ is consistently increasing. At $\Delta/\delta = 20$ the protein configuration is trapped in a high $U$ state.

An additional related problem in the energy minimization of this bonded protein is illustrated in the next Fig. S4. Large negative values of $U$ arise during minimization in some starting time points across conventional MD, which is an artefact that requires adjustment of Gromacs minimization parameters. Such adjustment is also required when $U$ is large positive, since then the subsequent NVE stage takes too much time to minimize the $U$ of the system and there is no speedup in the given virtual MD step compared to conventional MD (Fig. S5a); sometimes $U$ even does not converge.

It is found that the best way to proceed with energy minimization is using MD NVE simulation (for parameters, see Table S1). The NVE run is performed in short segments, with setting atomic velocities to zero before starting next segment. In each of these NVE segments the $U$ first decreases, and accordingly the atomic kinetic energy increases which forces atoms to get in closer contact with each other and also stretching the bonds that leads to eventual increase in $U$, seen as spikes on $U(t)$ plot (Fig. 5 and S5). Frequent cooling of the system helps to remove this excess motion and thereby encourages atoms to move in the direction of lower $U$. In addition, MD NVE is faster than NVT or NPT due to the lack of temperature and pressure couplings, which helps to increase speedup in virtual MD. As the system evolves in time and gradually equilibrates in conventional MD, the NVE stage can be made shorter due to lower $U$ values in the initial MD ensemble, as well as in the PA and EM configurations.

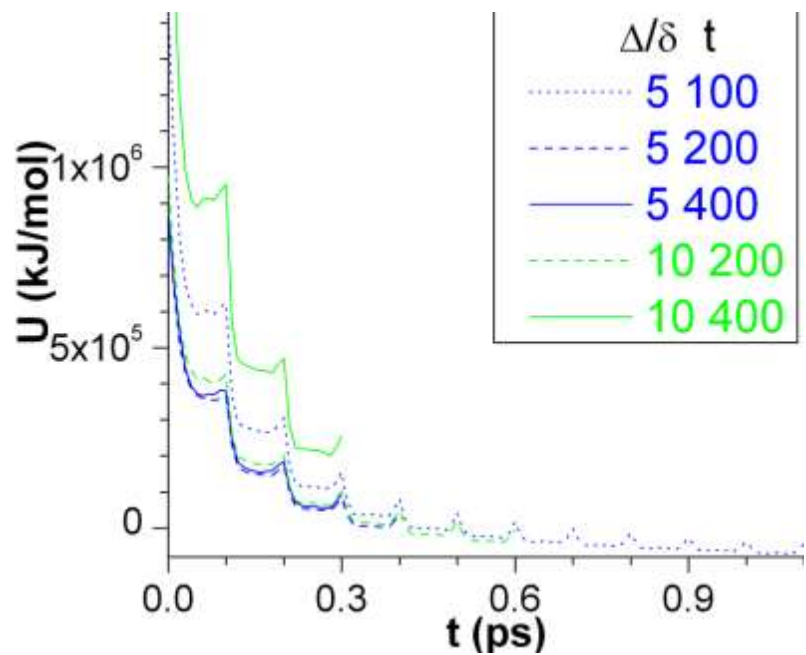

Fig. 5. Potential energy of pertussis toxin during NVE energy minimization. Values of $\Delta/\delta$ and PA-extrapolated time are shown in the legend. The case of large ratio $\Delta/\delta = 20$ is shown in Fig. S5.

At larger ratio $\Delta/\delta = 20$, the MD NVE-assisted energy minimization with conventional MD parameters fails to converge to a low-energy configuration (Fig. S5a). After adjustment of Gromacs minimization cut-off parameters, the NVE minimization proceeds successfully at large $\Delta/\delta = 20$ (Fig. S5b). Thermal equilibration of the NVE configuration is illustrated in Fig. S6. Validation of the resulting EQ configuration (the final configuration of the first virtual MD step) is presented in Fig. S7.

Considerable speedup over traditional MD, $s(\Delta/\delta)$, has been achieved in the PA-based virtual MD for the 1PRT structural transition simulation: the average values across tested all-atom configurations are $s(5) \approx 3.8$, $s(10) \approx 7.6$, $s(20) \approx 14.8$, and the maximum values are $s(5) = 4.5$, $s(10) = 9.0$, $s(20) = 17.3$. Fig. 6 shows the speedup versus PA-extrapolated time $t$ for three values of $\Delta/\delta$. In measuring $s$, input/output time in minimization and equilibration stages was accounted for.

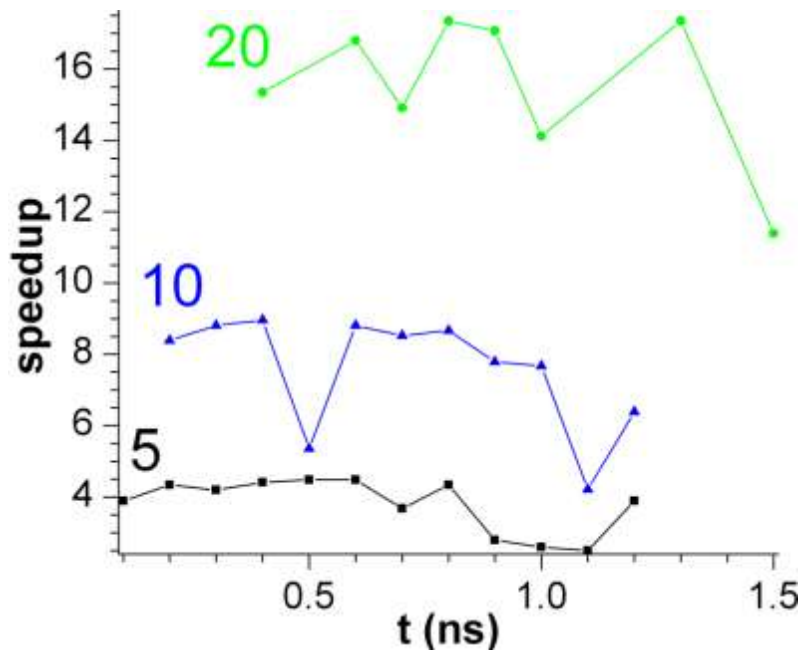

Fig. 6. Speedup of virtual versus conventional MD for 1PRT protein simulation as a function of the time $t$ for three ratios of $\Delta/\delta$ (shown in different *colors*).

The characteristic time of pertussis toxin protein evolution increases as the system approaches equilibrium (Fig. 7), (and similarly the potential energy decreases, Fig. S1). The characteristic time is assessed by the ratio of first to second time derivative of the PA **Error! Reference source not found.** at the end of each extrapolation interval, $\tau = 2\Delta$, averaged over all atoms. Its values are similar for the three Cartesian components of atom displacements, as well as for different $\Delta/\delta$ ratios, which confirms consistency of the chosen PA extrapolation scheme with respect to its parameters. The characteristic time of 1PRT evolution is well above the maximum PA step $\Delta = 20$ ps, so that the $\Delta/\delta$ ratio can be considerably increased. This implies a potential for further increase in virtual MD speedup for this average-size protein.

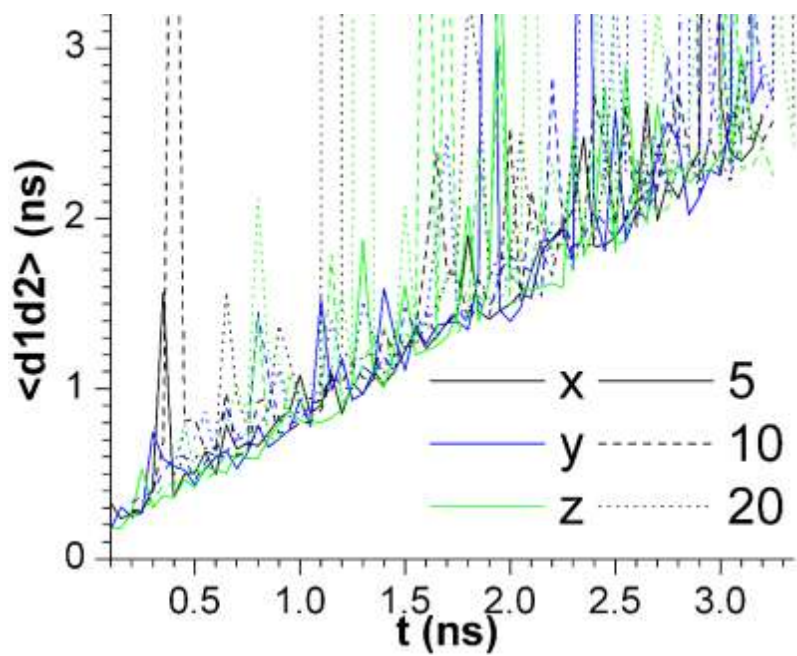

Fig. 7. Time dependence of the characteristic time of pertussis toxin evolution. Different *line styles* correspond to $\Delta/\delta = 5, 10, 20$ with $\delta = 10$ ps, while *colors* distinguish between Cartesian components.

To increase the virtual MD speedup, an attempt is made to improve quality of energy minimization that would allow to shorten the equilibration stage. A preliminary assessment of the Microstate Sparse Reconstruction scheme (MSR)[32] in minimizing the energy of the PA configurations shows that MSR is applicable to average-size proteins such as 1PRT. Fig. S8 shows a reasonable decrease in the potential energy of 1PRT in course of MSR iterations for three PA configurations generated at different stages of conventional MD. Analysis suggests that the impact of the MSR stage on subsequent long-time MD run is not sensitive to the number of MSR iterations (Fig. S9), at least if the energy minimization stage starts with MSR, or consists only of MSR. This is an advantage in real discrete evolution simulations where a sequence of many CG virtual MD steps has to be performed. Speedup in virtual MD could be improved by shortening the minimization and equilibration stages using MSR. However, this would require a more detailed analysis of MSR, such as an optimization of the number of iterations and finding the best way of combining the MSR with conventional minimization.

## HPV16 capsid in vacuum

*Conventional energy minimization*

As expected, the PA configurations of HPV16 T=1 capsid-like structure have some number of steric conflicts which have to be resolved using energy minimization. First a conventional energy minimization scheme implemented in Gromacs is considered. It combines conjugate gradient and steep descent methods, with the default choice of one conjugate gradient step per 10 steep descent steps. It is shown that 50000 minimization steps is sufficient to enable subsequent MD simulation in Gromacs (e.g., NVT simulation, Fig. 8) without having to perform thermal equilibration.

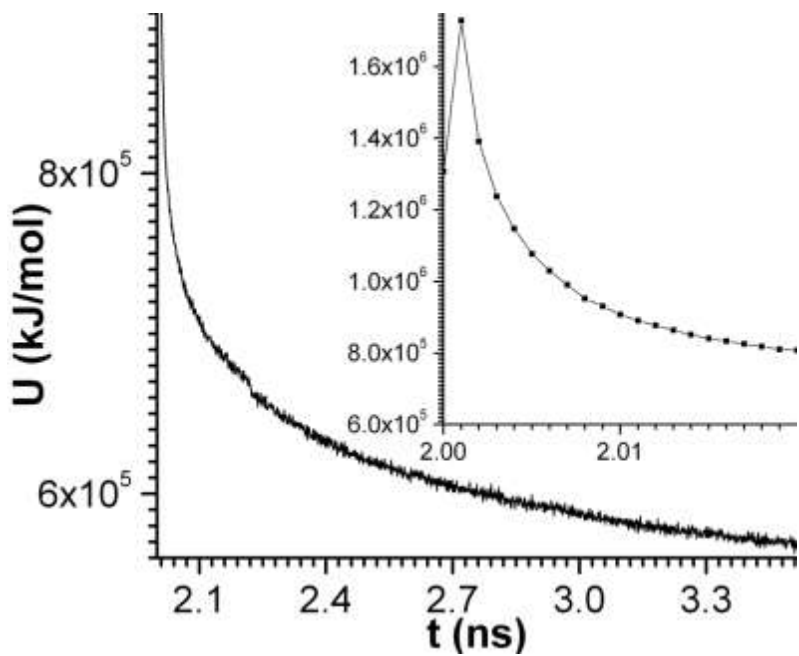

Fig. 8. Validation of the EM configuration of HPV16 T=1 capsid-like structure in vacuum (obtained from the PA configuration after 50000 conventional energy minimization steps) by MD NVT simulation (1$^{st}$ stage of the 2$^{nd}$ virtual MD step). $\Delta/\delta = 5$, $t - 2\Delta = 0$.

However, the potential energy in this NVT simulation is much higher than that in the conventional MD and four other ensemble MD simulations starting from the same initial coordinates (Fig. 9).

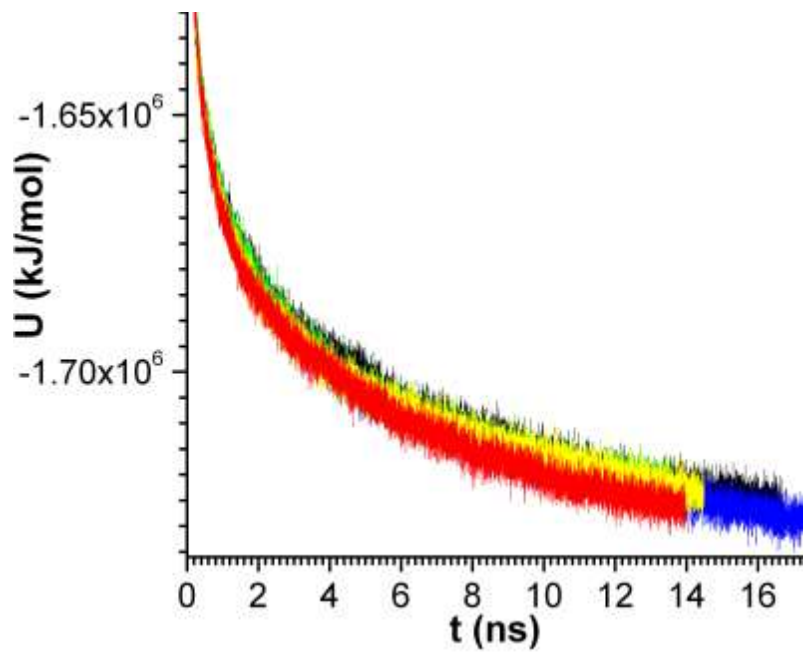

Fig. 9. Potential energy of HPV16 T=1 capsid-like structure in vacuum in the course of its five ensemble MD NVT simulations. All simulations were launched from the same structure – the initial configuration in the conventional MD NVT simulation – but with different initial velocities randomly generated using Maxwell distribution at 300 K. Slight differences in the energy between ensemble MD simulations reflect orbital instability of the conventional MD.

During 3.5 ns of NVT run, the EM configuration undergoes a compactification (Fig. 10), which exceeds that one in ensemble MD simulations (Fig. 11).

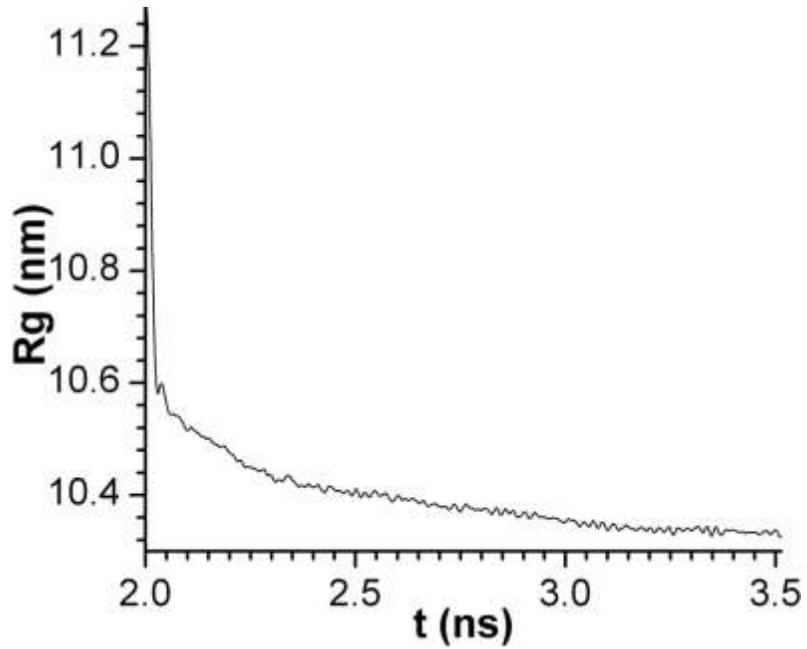

Fig. 10. Radius of gyration of the EM configuration in the course of its MD NVT simulation. $\Delta/\delta =10$, $t-2\Delta =2$ ns.

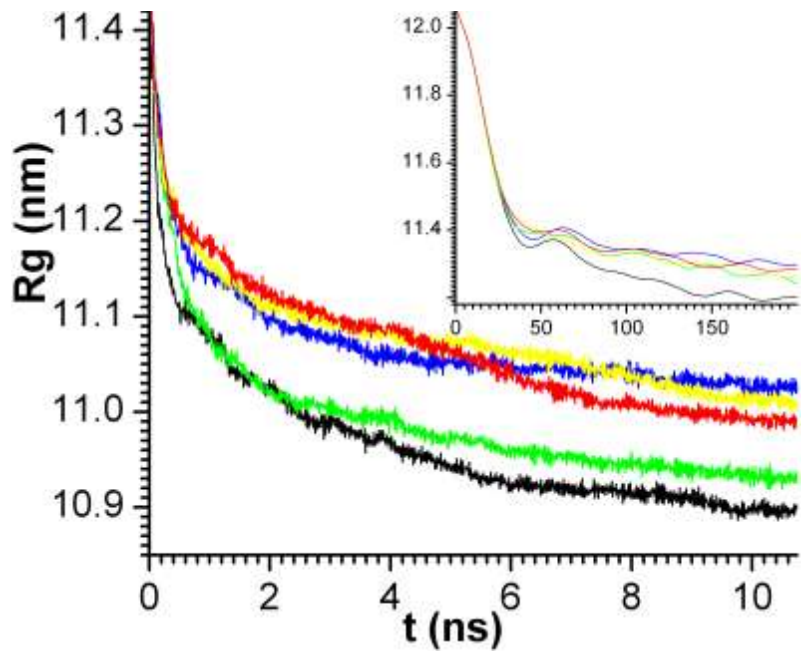

Fig. 11. Radius of gyration in ensemble of five MD NVT simulations of HPV16 T=1 capsid-like structure in vacuum.

Together with the observation of higher potential energy in virtual MD compared to conventional MD, this result indicates that the system is trapped in a high local energy minimum. Therefore, an improved energy minimization protocol is developed below.

*Conventional energy minimization combined with MSR.*

The utility of MSR in minimizing the energy of large systems has been studied by applying it to the HPV16 T=1 capsid-like structure in vacuum. When applied directly to the PA configurations (as suggested by its above mentioned successful application to the smaller 1PRT system), MSR frequently fails (its iterations diverge). Therefore, some amount of conventional energy minimization has to be performed before starting MSR for most of the PA conformations. The effect of the amount of this preliminary conventional minimization on the performance of combined conventional – MSR energy minimization scheme is illustrated in Fig. 12. It does not show a substantial decrease in the potential energy for the resulting configurations, suggesting the necessity of a detailed study of the effect of duration of MSR (i.e., the number of MSR iterations) on the quality of minimization (see below).

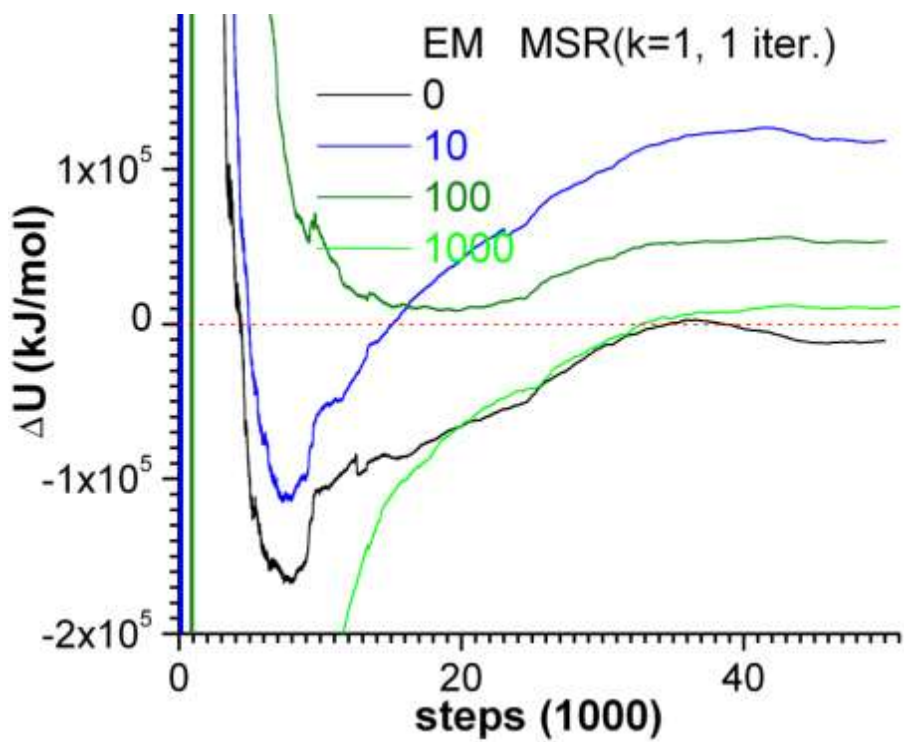

Fig. 12. Potential energy in course of the combined conventional – MSR energy minimization relative to that for conventional minimization, shown for different amounts of conventional

minimization performed prior to MSR (as in the *legend*) to determine if there is an optimum. One iteration with 4 CG variables[32] is used in the MSR calculation. Very deep minima at small numbers of minimization steps (below 1000) are due to the large magnitude of the energies in the beginning of minimization, and are not indicative of any improvement in the minimization protocol. After approximately 1000 minimization steps, these minima change to very high maxima, followed by a local minimum at 8-20 thousand steps which can be both above and below zero for some of the durations of initial conventional energy minimization. For the given PA configuration, a modest improvement over the conventional minimization was achieved only when starting MSR immediately from the PA configuration.

Effect of the number of MSR iterations on the potential energy after combined conventional – MSR minimization was studied. Only minor improvement in the potential energy is achieved, and without clear dependence on the number of MSR iterations (Fig. 13). Difference in the potential energy between combined and conventional schemes was small (and required logarithmic scale for graphical purposes).

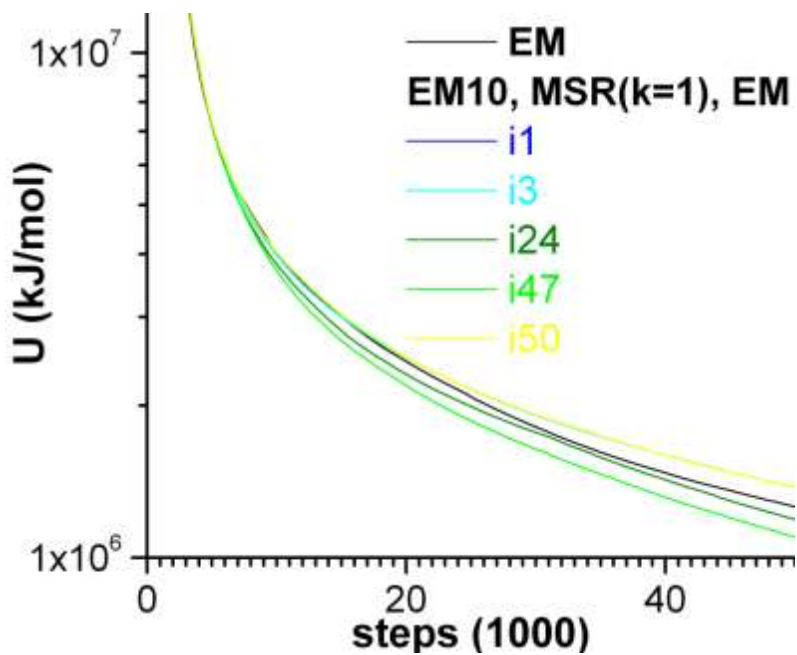

Fig. 13. Potential energy (in logarithmic scale) for the most promising sets of MSR parameters: preliminary energy minimization is performed for 10 steps, MSR used 4 CG variables. For 24 and 47 MSR iterations the final energy is slightly lower than that in conventional minimization, while for increased number of 50 iterations it again becomes higher, suggesting that there is no clear dependence of the potential energy of final minimized state on the number of MSR iterations.

*Additional energy minimization using MD NVE.*

The problem of PA configuration trapping in a local energy minimum is addressed below by adding energy minimization via MD NVE simulation. The NVE is performed in short segments with discarding the system kinetic energy before next segment. This scheme removes excess potential energy which converts into kinetic energy during each NVE segment. The NVE minimization also suggests the possibility of increasing speedup in virtual MD from shortening the thermal equilibration stage.

Starting with the PA configuration of HPV16 T=1 capsid-like structure in vacuum for $\Delta/\delta = 10$ minimized for 50,000 steps in Gromacs, multiple MD NVE runs were launched with different segment durations *L*, i.e., different periods between setting atomic velocities to zero. The segment duration was kept constant throughout each NVE minimization. The timecourses of potential energy for different *L* were compared with that in conventional energy minimization with Gromacs (Fig. 14). It is seen that too short ($L = 1$ and 3 fs) NVE intervals minimize energy slower than the conventional combination of steep descent and conjugate gradient methods (1 conjugate gradient step per 10 steep descent steps), and that the NVE simulation quickly fails after strong increase in the potential energy (see insert in Fig. 14). Increasing the period *L*, one makes the minimization process much more efficient than conventional minimization. The optimal segment duration in the beginning of NVE minimization is $L = 3$ *ps* for the studied system. Too long NVE segments require too long time to reach energy minimum due to saturation of the potential to kinetic energy conversion within each segment.

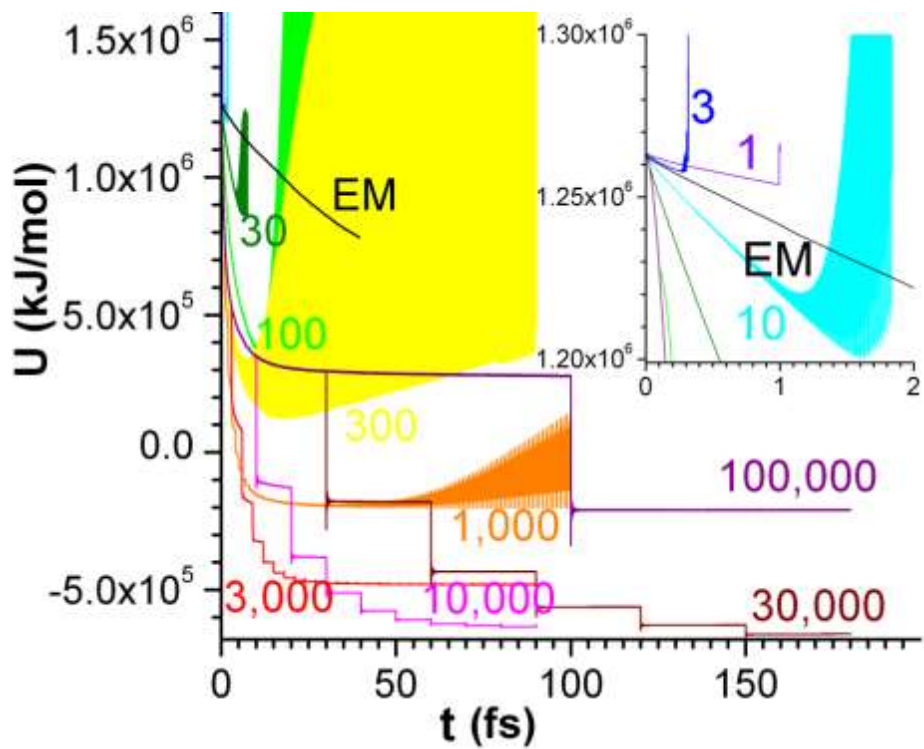

Fig. 14. Potential energy of HPV16 T=1 capsid-like structure in vacuum during MD NVE energy minimization. $\Delta/\delta = 10$, $t = 2$ ns. (*Black*) Conventional minimization in Gromacs using combination of steep descent and conjugate gradient methods. (*Color*) Minimization using MD NVE with periodic discarding of the kinetic energy (periods are shown in femtoseconds by *color labels*).

    Other observations on the parameter $L$ are as follows. For $L \leq 30$ *fs* the MD NVE simulation invariably crashes, while it is typically longer for larger $L$. After reaching an energy minimum for a given $L$, the system starts increasing both the lowest and highest energy values within each NVE segment. The same occurs at $100\ fs \leq L \leq 300\ fs$, but without breaking the simulation, at least within the sampled overall durations of NVE simulations. The simulation time becomes too long, and the potential energy keeps growing, which makes continuation of NVE for such intervals non-prospective. At $L = 1000\ fs$ the lowest potential energy value within an interval stays same. Starting at $L = 3000\ fs$ the maximum potential energy value within an interval does not appreciably exceed the minimum energy value, and the NVE minimization seems to be stable, i.e., converging to a local energy minimum. Larger $L$ lead to lower potential energy values. However, the energy is still considerably higher (Fig. 14) than $-1.68259 \cdot 10^6$

kJ/mol in conventional MD at the same projected time $t = 2$ ns. Reduction of the remaining energy gap between EM and conventional MD configurations may be addressed by further optimization of NVE minimization parameters.

The final NVE energy-minimized configuration for each $L$ is compared to those taken throughout conventional MD simulation in vacuum (Fig. 15). RMSD curves for $L = 1$, 3, and 10 *fs* are almost identical. At increasing $L$, the wide and shallow minimum of RMSD turns into a clearly defined minimum. Most striking is the location of this minimum of RMSD at exactly the projected time 2 ns for moderately large NVE intervals ($L = 1000$ and $3000$ *fs*). This suggests a possibility of retaining realistic time scale of the dynamics with virtual MD. As $L$ increases further, the resulting NVE configuration starts resembling MD structures at later times more than those immediately prior to the projected time. This, together with the increase in RMSD at increasing $L$, reflects the tendency of prolonged NVE minimization to impart further evolution to the system.

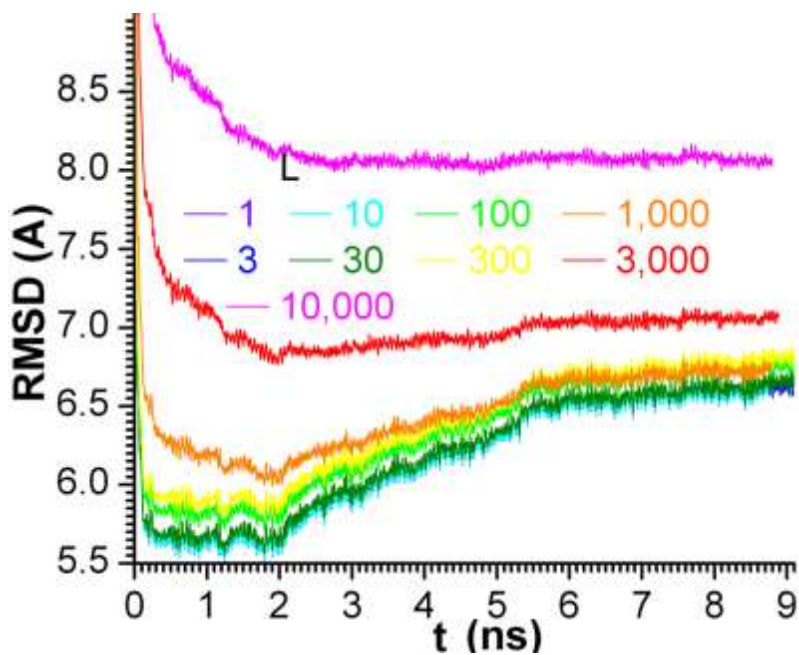

Fig. 15. RMSD of HPV16 T=1 capsid-like structure backbone in NVE EM configurations for different durations $L$ (in fs, see *legend*) from the conventional MD configuration at corresponding time.

This additional evolution of the system imparted by the NVE minimization is more clearly illustrated by the size-related quantity – the radius of gyration, $R_g$ (Fig. 16). $R_g$ decreases in course of NVE simulation and as $L$ increases. Starting at $L = 1$ ps, the final structure becomes more compact than in conventional MD (where $R_g = 11.0173$ nm at 2 ns, Fig. 11). Starting at $L = 30$ ps, the compactification saturates. Insert in Fig. 11 shows that for some short intervals $L$ the structure starts expanding before corresponding NVE simulation breaks.

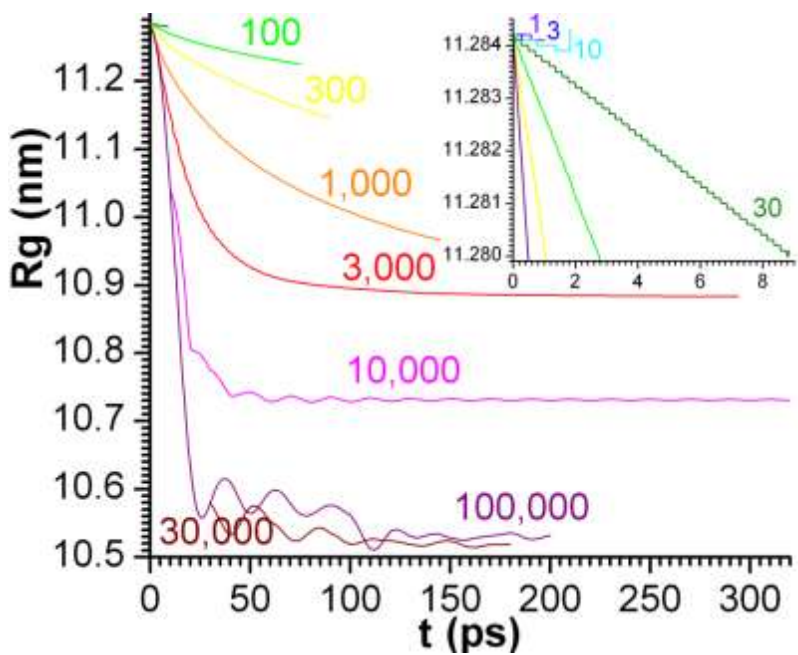

Fig. 16. Radius of gyration of HPV16 T=1 capsid-like structure during NVE energy minimization showing compactification beyond the size of the structure in conventional MD.

A good quality of the EM configuration obtained by the combined conventional – NVE minimizations of the PA configuration can be seen from its comparison to the ensemble MD simulations launched from the same original structure at time $t - 2\Delta = 0$, but with different initial velocities. The NVE configuration has size and RMSD relative to conventional MD simulation (which is the first of five ensemble MD simulations) that are within the range of those for the members of the MD ensemble. In ensemble MD, the capsid RMSD at $t = 2$ ns ranges from 7.087 to 7.740 Å for the second through fifth MD simulations relative to the first one due to orbital instability (Fig. 17), while for the NVE EM configuration it is 7.540 A for $L = 3$ ps (this value is recommended for the initial stage of NVE energy minimization, see Fig. 15).

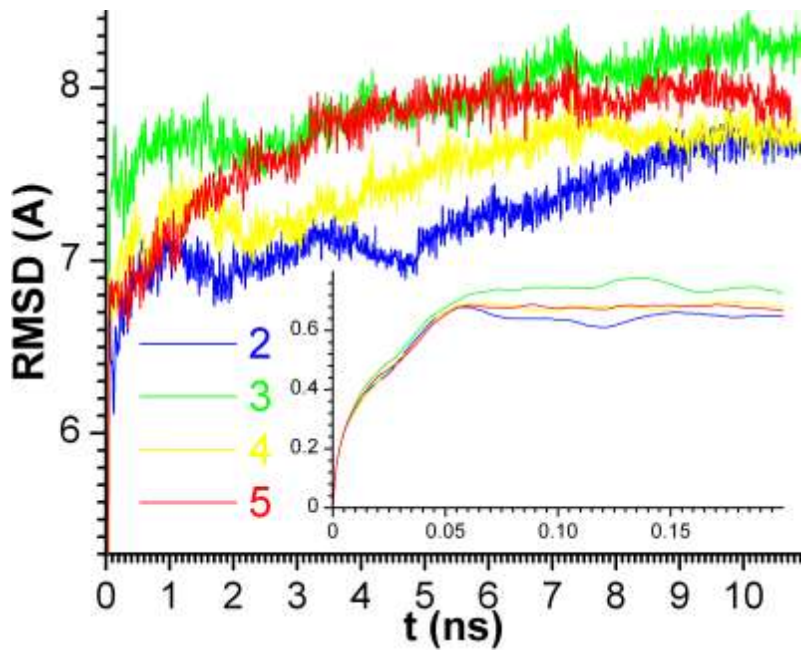

Fig. 17. RMSD of the second through fifth ensemble MD simulations relative to the first one reveals the orbital instability of MD.

Radius of gyration also agrees well with ensemble MD. The radius is between 110.173 and 111.281 Å in the five ensemble MD simulations (Fig. 11). For the NVE EM configuration, the closest agreement is achieved at $L = 1$ ps for which the radius is 109.668 Å. At $L = 3$ ps one has lower value 108.836 Å. As $L$ is increased, the structure becomes more compact and reaches 105.260 Å radius at $L = 100$ ps (Fig. 16). This suggests that the optimal duration of NVE simulation segments is $L = 1$ ps.

**Argon Droplet Coalescence**

The non-bonded argon system has liquid-like behavior with atom fluctuations (Fig. 2a) larger than in the bonded protein systems considered above (Fig. 2b). This results in the lower quality of the PA configurations compared to those for the bonded systems, as is manifested in the higher potential energies $U$ of the PA configurations (per atom) of the coalescing argon nanodroplets (Fig. S10). During conventional energy minimization of the PA configurations as part of the virtual MD algorithm (Fig. 1) $U$ decreases rapidly (Fig. S12), and the system configuration is seldom trapped in a local energy minimum for the studied initial MD configurations of this non-bonded system. A sufficient (and *apriori* unknown) degree of energy minimization is required for numerical stability of the next stage of thermal equilibration during updating of the system to the next CG timestep. The $U$ of the EM configurations for large enough number of energy minimization steps (Fig. S13) is considerably lower than that of conventional MD (Fig. S10) (due to the absence of thermal motion in the energy minimization stage of virtual MD). An increase in the minimized $U$ is observed as the $\Delta/\delta$ ratio is increased (Figs. S12 and S13). Note that MSR cannot be applied to a non-bonded system because it only optimizes bonded parameters such as interatomic bond and angle distances.

For this non-bonded system, it is possible to continue the virtual MD simulation immediately after energy minimization, i.e., without thermal equilibration (Fig. S14). However, in the subsequent MD NVT run $U$ decreases (when a small number of minimization steps is performed) or increases (for a larger number of minimization steps) and eventually equilibrates closer to its value in conventional MD (see Supplementary Information). For a small number of minimization steps (100 and 250), the $U$ equilibration may take 9 to 18 ps of NVT simulation (Table S2), making speedup in virtual MD impossible. For large enough number of minimization steps (1000), $U$ equilibrates within 1.1 ps, but temperature equilibration occurs later, typically after 1.6 ps. This duration of NVT equilibration, short relative to $\Delta$, implies that the computation can be very efficient.

In order to validate the equilibrated configurations, a next virtual MD step is started from them by extending the duration of MD NVT simulation for a duration of $\delta$ and longer (up to 100 ps starting from the beginning of equilibration stage). The microstate obtained after sufficiently long MD NVT simulation is robust with respect to change in the number of energy minimization steps. Effect of increased degree of energy minimization on the subsequent MD run is studied by taking 100, 250, and 1000 minimization steps (Figs. S14 and S15) for $\Delta/\delta = \{5, 10, 20\}$. The $U$ converges to the same value for different number of minimization steps; this value grows slightly as $\Delta/\delta$ increases. At shorter MD time intervals, e.g., 5 ps required for subsequent virtual MD steps, the higher degree of energy minimization gives lower $U$ after thermal equilibration (Fig. S15).

RMSD between conventional MD and PA configurations at the same extrapolated time (Fig. 18) is very large due to the liquid-like character of the motion of non-bonded atoms. It grows strongly after 0.9 ns due to system attraction to its mirror images in conventional MD (due to periodic boundary condition artefacts; a larger simulation box would avoid this, but would not affect the estimation of virtual MD efficiency). However, this RMSD is considerably smaller than system size and the RMSD of both the MD and PA with respect to the initial system state from which conventional MD has started (Fig. S16). In addition, the latter RMSD values for PA configurations are very close to those for conventional MD (Fig. S16), which suggests that the PA extrapolation closely follows the system evolution trend in conventional MD. Even for the largest ratio $\Delta/\delta$ =20 the basic physics is quantitatively characterized by the PA.

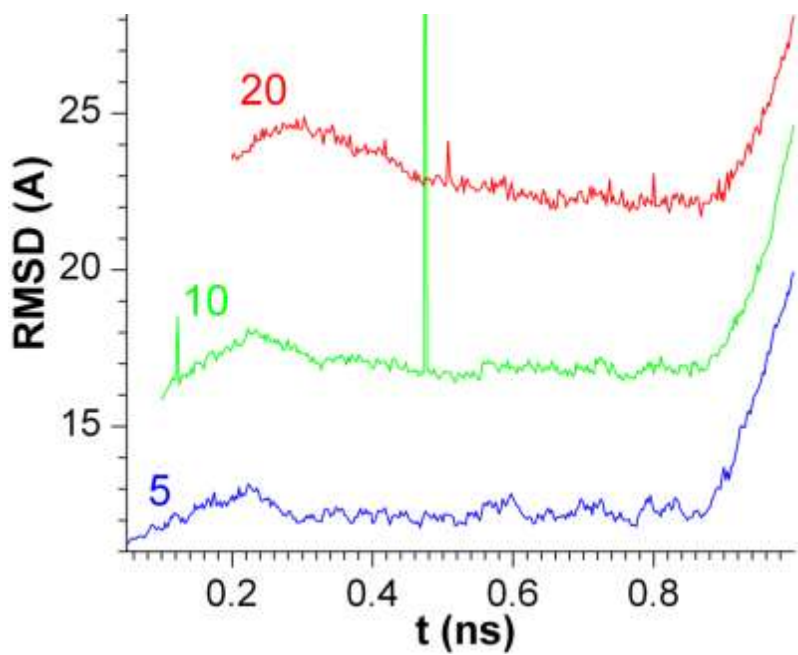

Fig. 18. RMSD between MD and PA configurations at the same extrapolated time in course of system evolution, and it's growth with the increase of $\Delta/\delta$ (*labeled* in the plot).

Radius of gyration for the PA configurations follows its trend in conventional MD (Fig. 19). As expected, the difference between PA and MD grows with increase of $\Delta/\delta$, and decreases with increasing the number of energy minimization steps. Likewise, the relative difference between MD and NVT thermal equilibration radii becomes smaller with increased number of minimization steps (Fig. 20).

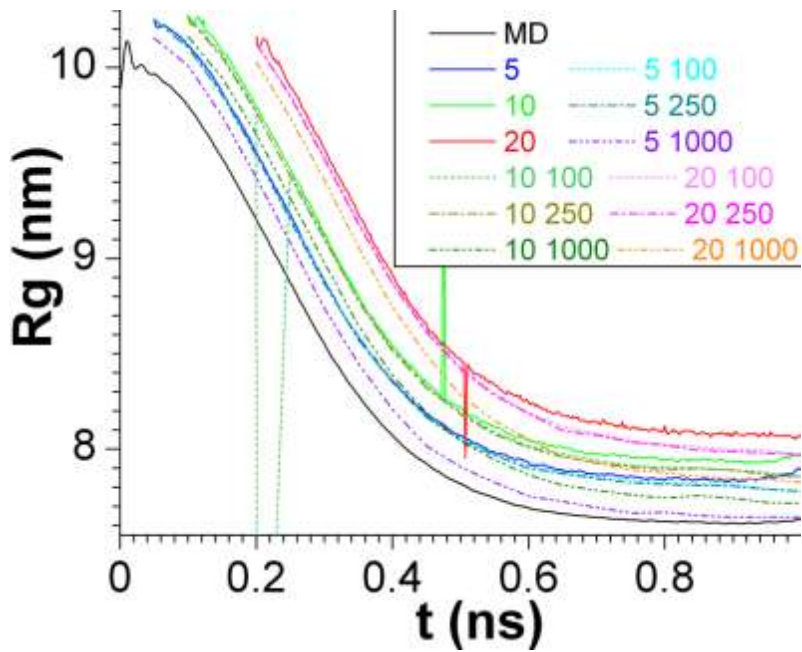

Fig. 19. Radius of gyration for MD (*black*), PA (*color*, *solid* curves), and EM (*color*, *non-solid* curves) configurations versus evolution time for different $\Delta/\delta$ (distinguished by *colors*, see *first entry* in the *legend* for each curve) and numbers of minimization steps (distinguished by *line styles*, see *second entry* in the *legend*).

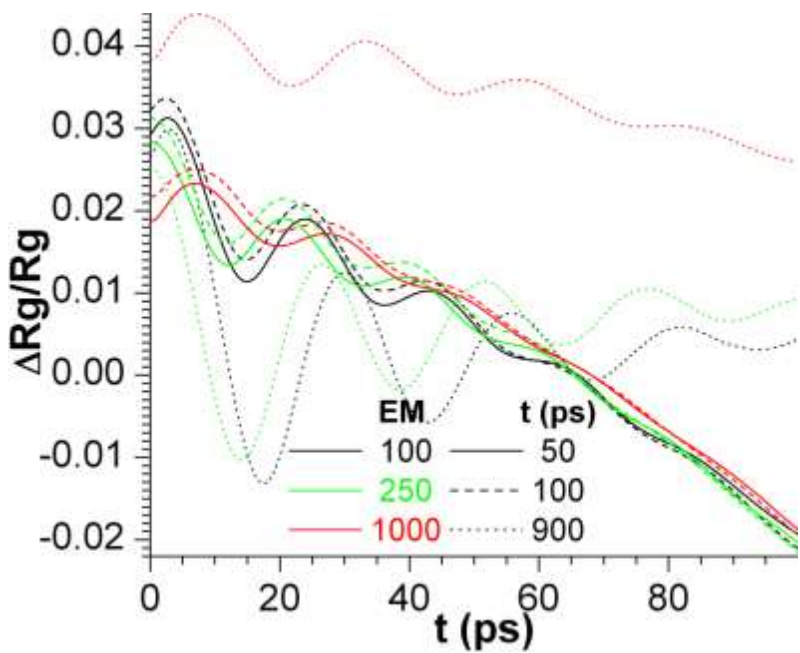

Fig. 20. Error of the radius of gyration for NVT configurations relative to conventional MD for different numbers of energy minimization steps (different *colors*, see *legend*) and evolution time (different *line styles*). $\Delta/\delta = 5$. Except for 1000 minimization steps at t = 900 ps, increase in the number of minimization steps leads to a smaller error.

Analysis of the particle density shows that the core of the coalesced argon droplets in the PA configurations is very similar to that in MD simulation (Fig. 21). Noticeable differences start appearing at $\Delta/\delta = 20$. This suggests that the $\Delta/\delta$ ratio should be limited by 20 to preserve accuracy and stability for the present problem, and likely for other non-bonded systems. Considerable difference in the radius of gyration between PA and MD configurations (Fig. 19) only indicates larger spread of a small fraction of atoms outside the core in the PA configuration.

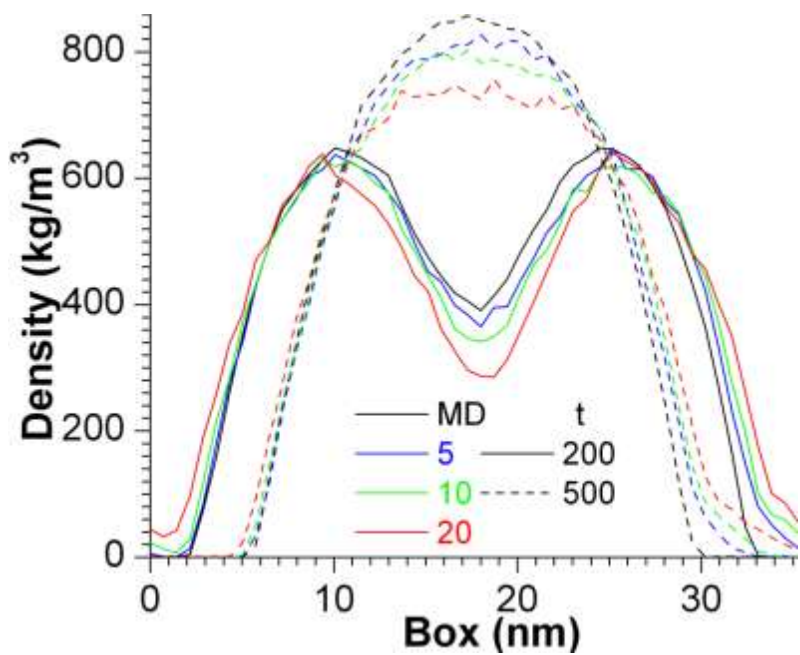

Fig. 21. Partial density along axis connecting initial positions of the centers of two coalescing argon droplets in MD (*black*) and PA configurations (*color*; $\Delta/\delta$ ratio is shown in the *legend*) at different evolution times: 200 ps (*solid*) and 500 ps (*dash*).

Due to the indistinguishability of argon atoms and the instability of Newtonian orbits, it does not make sense to distinguish between particles in this non-bonded system. Therefore, RMSD between MD and PA configurations is considered without accounting for particle labels.

This implies the need to pair atoms in a reference and a considered configuration which are closest to each other. The algorithm for achieving this used in this study is as follows. Each particle $i = 1, \ldots N$ in the considered configuration at time $t$ is paired one-by-one with particle $j$ in the reference configuration at the same time which is the closest to particle $i$ and is not yet paired. A sum of squared distances between atoms in each $i$-$j$ pair is calculated, divided by $N$, and the square root is taken. This RMSD between the MD and PA configurations at the same extrapolated time (Fig. 22) is considerably smaller for all considered $\Delta/\delta$ values than the one that does not account for particle indistinguishability (Fig. 18). This suggests that accounting for particle indistinguishability yields better comparison metric. The indistinguishability-based RMSD relative to the initial MD configuration shows that the PA configurations closely follow the evolution trend in MD (Fig. S17), as is the case for the RMSD with account for particle labels in Fig. S16.

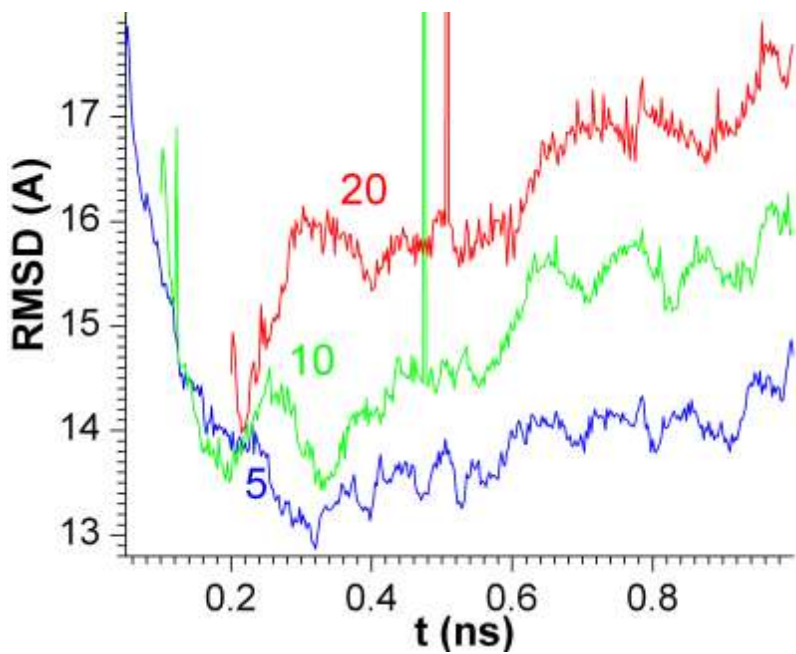

Fig. 22. RMSD based on minimal distance pairing, to address particle indistinguishability, between PA and conventional MD configuration at corresponding extrapolated time for argon droplets. Three $\Delta/\delta$ values are compared with each other (*labeled* in the plot).

The speedup of the virtual MD relative to conventional MD, $s(\Delta/\delta)$, for the non-bonded systems like argon droplets is expected to be greater than that for the bonded systems studied above because less minimization and equilibration steps are required for the non-bonded systems. Indeed, PA for argon system only required 100 energy minimization steps and no thermal equilibration was necessary to launch MD simulation from the resulting configurations. In absence of thermalization, s(5) = 4.66, which is close to the theoretical limit $\Delta/\delta = 5$. If 1.6 ps of NVT equilibration would be performed (for the purpose of thermal equilibration, before energy converges for the small number of minimization steps), $s$ = 3.59. Even if 11 ps of NVT would be performed (recommended for energy convergence, see Table S2), still some acceleration $s$ = 1.53 would be achieved. In case if maximum number of 1000 minimization steps is performed, in absence of thermal equilibration s = 3.69, and with the recommended 1.6 ps of equilibration s = 2.98. The above measurements were performed for $\Delta/\delta = 5$ at extrapolated time 50 ps of the argon droplets MD simulation. For $\Delta/\delta$ =10, the maximum speedup is 9.32, achieved with 100 minimization steps and no thermal equilibration. With 1.6 ps thermal equilibration, s = 7.18; with 11 ps equilibration, s = 3.06. With 1000 minimization steps and no equilibration s = 7.37; with 1.6 ps equilibration s = 5.97. Hence a 6-fold speedup with retaining good accuracy of virtual MD simulation is expected when using $\Delta/\delta$ =10 in the PA extrapolation stage (with reasonable assumption that the thermal equilibration time is close to that for $\Delta/\delta = 5$). For $\Delta/\delta = 20$, the maximum speedup is 18.64. With 100 minimization steps and 1.6 ps equilibration s = 14.36, with 11 ps equilibration it is 6.11. With 1000 minimization steps and no thermal equilibration s = 14.75, and with 1.6 ps equilibration s = 11.93. This yields 12-fold speedup with retaining good accuracy. Underlying benchmarks for each of the MD stages of virtual MD are presented in the Supplementary Information (Figs. S19 – S20).

**IV. Discussion and Conclusion**

The proposed virtual MD method based on extrapolation of atomic coordinates with Pade approximants is demonstrated to be promising as an efficient atom-resolved MD algorithm that accurately simulates the long-time evolution of bonded and nonbonded nanosystems. In essence, the method is a multiscale Markov model wherein transition probabilities are constructed using a conventional short MD simulation with infrequent Markov transitions. It does not require introduction of collective variables or reaction coordinates, or calibration of energy-related system parameters and system equilibration, unlike in other multiscale MD or accelerated MD methods.

The dynamics of $N$ atoms is modeled here via a set of $3N$ virtual atomic coordinates. The challenge of virtual MD is to construct the transition probability for the atoms, i.e., to predict the likely atomic position from that at time $t = (n-1)\Delta$ (and earlier) to that at $t = n\Delta$. An explicit semi-analytical PA scheme is developed that incorporates notions of the Ito formula for stochastic processes.[18] Information from earlier all-atom configurations obtained by short conventional MD runs is used to evolve atoms over long time scales and thereby coherent phenomena can be captured. Thus, in virtual MD a temporal coarse-graining is used that only indirectly accounts for the $10^{-14}$ second atomic fluctuations induced by atomic vibrations/collisions. By tracking the omnipresent fluctuations via the Ito formula, virtual MD captures frictional effects, collective oscillations, and other mesoscopic phenomena.

Assessment of the virtual MD was performed by taking historical information at multiple time points of the conventional MD simulations for three nanosystems of different size and interatomic bonding character. Corresponding all-atom states were evolved over one virtual MD step, i.e., were subjected to PA extrapolation, energy minimization and, in most cases, thermal equilibration. Resulting virtual MD microstates were compared to the MD microstates at corresponding times. Selected virtual MD microstates were used to launch a short MD simulation stage of the second virtual MD step to validate the proposed approach.

A main advantage of the PA method is that it restricts extrapolation values at large times unlike, e.g., Taylor series expansions. This facilitates extrapolation of each of the 3N atomic coordinates to the next CG timestep (which is much greater than the MD timestep). In this study, linear asymptotics of the PA in time is shown to work well for the considered systems and timestep sizes. Theoretical efficiency of the implied numerical scheme is $O(\Delta/\delta)$. That $\delta$ (the duration of conventional MD interval used to construct the PA) can be taken much less than $\Delta$ (the CG timestep) is justified via the stationarity feature of the atomistic fluctuations (whereby a many-atom system expresses a representative ensemble of microstates on a timescale shorter than that of overall system dynamics).[19]

Average speedup factor of 14.8 was achieved in the virtual MD step for pertussis toxin protein across different all-atom configurations, with maximum value 17.3 in one case. For HPV16 T=1 capsid-like structure in vacuum, the speedup is 1.56 for $\Delta/\delta = 5$. For the nonbonded argon droplets system at 70 K no equilibration was required to continue virtual MD trajectory, and the same amount of energy minimization could be used for all tested all-atom MD configurations. The resulting speedup is 2.98 for $\Delta/\delta = 5$, 5.97 for $\Delta/\delta = 10$, and 11.93 for $\Delta/\delta = 20$.

Efficiency of virtual MD is very sensitive to the quality of energy minimization via its effect on the number of necessary thermal equilibration steps, so that more advanced energy minimization schemes[33] should be tested. Virtual MD configurations could allow for atom overlapping or bond straining configurations, however, this could be avoided via constraint methods.[32]

Accuracy and speedup of virtual MD could be improved by using shorter intervals between historical MD configurations. To this end, it was shown that with 1 fs time interval between historical MD configurations and $\Delta = 10\delta = 100$ fs the virtual MD time advancement in most cases does not require minimization and equilibration. However, the MD interval $\delta$ should be at least as long as the stationarity time[19] to provide a reliable statistics on the atomic fluctuations. Estimation of the stationarity time requires frequent collection of the atomic coordinates on the order of femtosecond,[19] and thus was not attempted here.

We suggest that virtual MD will provide accelerated sampling of different states of molecules in a way similar to the accelerated MD[34] and flat histogram[35] techniques, but for much larger atomic and molecular systems. It is expected that the present method can be combined with other accelerated MD approaches to improve the efficiency of exploring the free energy landscape of biomolecules.

**ACKNOWLEDGMENTS**

The authors acknowledge the National Science Foundation (NSF: DMR-1533988 and CHE-1344263) and the Indiana Clinical and Translational Sciences Institute (GLUE initiative) for support. The authors also acknowledge support by the IU College of Arts and Sciences via the Center for Theoretical and Computational Nanoscience, and Lilly Endowment, Inc., through its support for the Indiana University Pervasive Technology Institute, and in part by the Indiana METACyt Initiative. The Indiana METACyt Initiative at IU is also supported in part by Lilly Endowment, Inc. Computations were performed on Karst supercomputer at Indiana University.

# Supplementary Information for

# Virtual Molecular Dynamics

Yuriy V. Sereda, Andrew Abi Mansour, Peter J. Ortoleva

**Pertussis Toxin Protein**

Potential energy of the pertussis toxin protein (PDB ID: 1PRT) in MD NVT simulation is shown in Fig. S1.

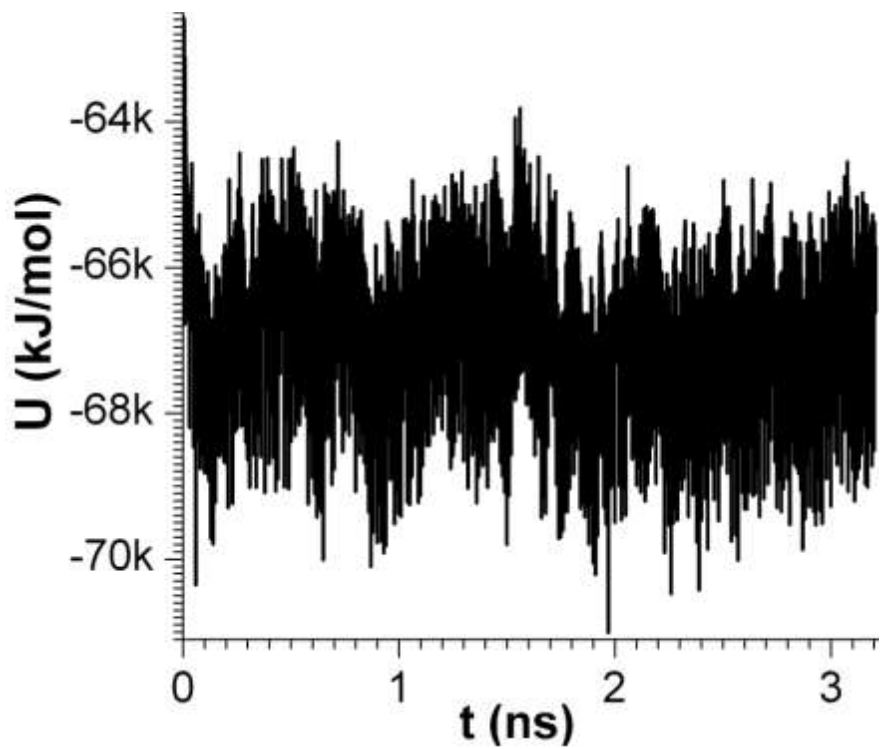

Fig. S1. Potential energy of pertussis toxin (PDB ID: 1PRT) in Gromacs MD NVT simulation.

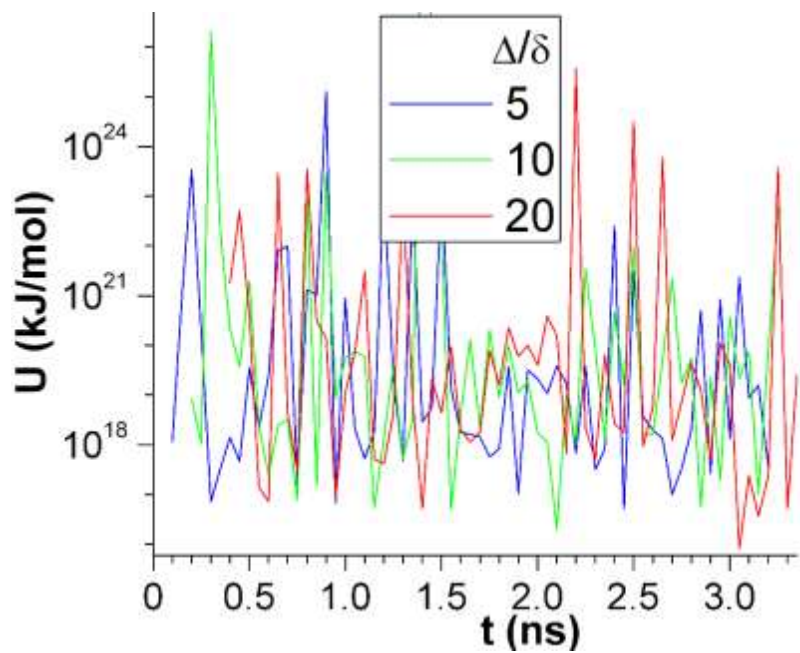

Fig. S2. Potential energy of 1PRT after PA extrapolation versus evolution time $t$.

Fig. S2 shows that $U(t)$ for the bonded protein system pertussis toxin after PA extrapolation is relatively low compared to that for the non-bonded argon nanodroplet system (Fig. S11).

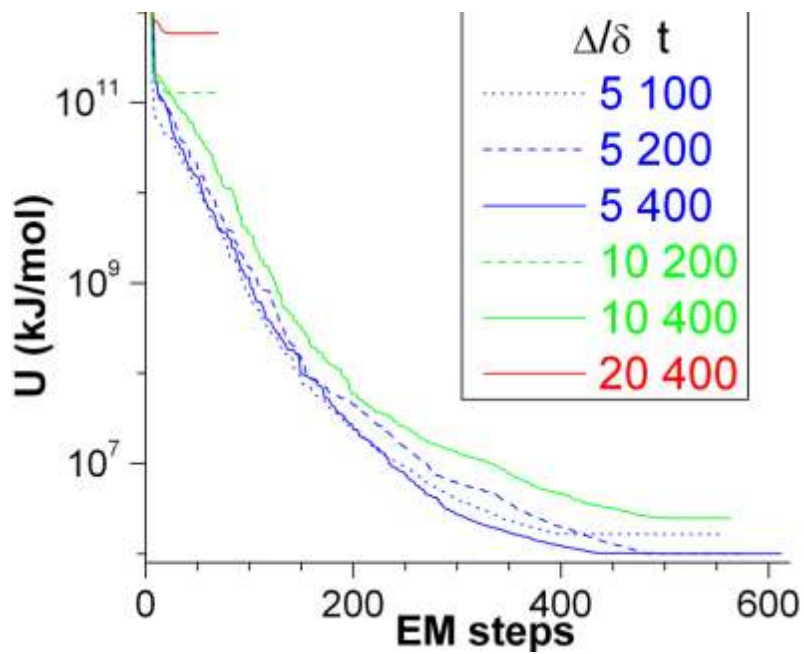

Fig. S3. Minimization of the potential energy of the PA-extrapolated atom configurations for 1PRT. *Colors* discriminate different $\Delta/\delta$ values in the PA, and *line styles* discriminate values of the system evolution time, as in the legend.

Table S1. Gromacs simulation parameters used in MD NVT and NPT, energy minimization, NVE minimization, and temperature equilibration stages of virtual MD to ensure numerical stability.

| **MD parameters** | 1PRT | HPV16 T=1 | Ar |
|---|---|---|---|
| *integrator* | md | md | md |
| *dt* [fs] | 2 | 1 | 0.5 |
| *pbc* | xyz | xyz | xyz |
| *constraints* | none | hbonds, none[*] | none |
| *nstcomm* | 1000 | 100 | 1 |
| *ns_type* | grid | grid | grid |
| *nstlist* | 1 | 10 | 1 |
| *rlist* [nm] | 1.0 | 1.4 | 1.2 |
| *cutoff-scheme* | group | group | group |
| *vdwtype* | cut-off | switch; rvdw-switch = 1.0 | cut-off; rvdw-switch = 0 |
| *rvdw* [nm] | 1.4 | 1.4 | 1.2 |
| *coulombtype* | PME | reaction-field; rcoulomb-switch = 0 | cut-off; rcoulomb-switch = 0 |
| *rcoulomb* [nm] | 1.0 | 1.4, 2.5[*] | 1.2 |
| *optimize_fft* | yes | N/A | N/A |
| *tcoupl* | v-rescale | v-rescale | v-rescale |
| *tau_t* [ps] | 0.1 | 0.1, 0.02[*] | 0.1 |
| *pcoupl* | Berendsen | no | no |
| *pcoupltype* | isotropic | N/A | N/A |
| *tau_p* [ps] | 0.5 | N/A | N/A |
| *DispCorr* | EnerPres | EnerPres | EnerPres |

[*] Parameters used for validation of EQ configurations (in a 2nd virtual MD step).

| EM parameters | 1PRT | HPV16 T=1 | Argon droplets |
|---|---|---|---|
| *integrator* | steep | steep | steep |
| *Pbc* | xyz | xyz | xyz |
| *emtol* | 100, 0.01 | 100 | 0.1 |
| *emstep* | 0.001, 0.01 | 0.01 | 0.01 |
| *constraints* | none | none | none |
| *nstcomm* | 1, 100 | 10 | 100 |
| *vdwtype* | cut-off | cut-off | cut-off |
| *coulombtype* | PME, reaction-field | reaction-field | cut-off; cutoff-scheme = group |
| *ns_type* | grid | grid | grid |
| *nstlist* | 10 | 10 | 1 |
| *rlist* [nm] | 1.4, 3 | 1.4 | 1.2 |
| *rcoulomb* [nm] | 1.4, 3 | 1.4 | N/A |
| *rvdw* [nm] | 1.4 | 1.4 | 1.2 |
| *DispCorr* | no | no | EnerPres |
| **NVE parameters**[**] | 1PRT | HPV16 T=1 | |
| *integrator* | md | md | |
| *dt* [fs] | 0.1 | 1 | |
| *Pbc* | no | xyz | |
| *constraints* | none | none | |
| *nstcomm* | 100 | 1000 | |
| *ns_type* | grid | grid | |
| *nstlist* | 10 | 10 | |
| *rlist* [nm] | 1.4, 3.0 | 1.4 | |
| *cutoff-scheme* | group | group | |
| *vdwtype* | cut-off | switch | |
| *rvdw-switch* [nm] | 1.3, 2.9 | 1.0 | |
| *rvdw* [nm] | 1.4, 3.0 | 1.4 | |
| *coulombtype* | reaction-field | reaction-field | |
| *rcoulomb* [nm] | 1.4, 3.0 | 1.4 | |
| *DispCorr* | no | EnerPres | |

**NVE was not required for non-bonded argon system.

| **EQ parameters**[***] | 1PRT | HPV16 T=1 | Ar |
|---|---|---|---|
| *integrator* | md | md | md |
| *dt* [fs] | 1 | 1 | 0.5 |
| *pbc* | xyz | xyz | xyz |
| *constraints* | all-bonds | none | |
| *nstcomm* | 100 | 100 | |
| *ns_type* | grid | grid | grid |
| *nstlist* | 10 | 1 | 1 |
| *rlist* [nm] | 3.0 | 1.4 | 1.2 |
| *cutoff-scheme* | group | group | |
| *vdwtype* | cut-off | switch; rvdw-switch = 1.0 | cut-off; rvdw-switch = 0 |
| *rvdw* [nm] | 3.0 | 1.4 | 1.2 |
| *Coulombtype* | PME | reaction-field | cut-off; rcoulomb-switch = 0 |
| *rcoulomb* [nm] | 3.0 | 1.4 | 1.2 |
| *fourierspacing* | 0.16 | 0.12 | |
| *optimize_fft* | no | N/A | |
| *tcoupl* | v-rescale | v-rescale | v-rescale |
| *tau_t* [ps] | 0.1 | 0.1 | 0.1 |
| *pcoupl* | Parrinello-Rahman | no | no |
| *pcoupltype* | isotropic | no | N/A |
| *tau_p* [ps] | 2.0 | N/A | N/A |
| *DispCorr* | no | EnerPres | EnerPres |

**Temperature equilibration of argon system was done as a part of MD NVT simulation, see "MD parameters" section of the Table S1.

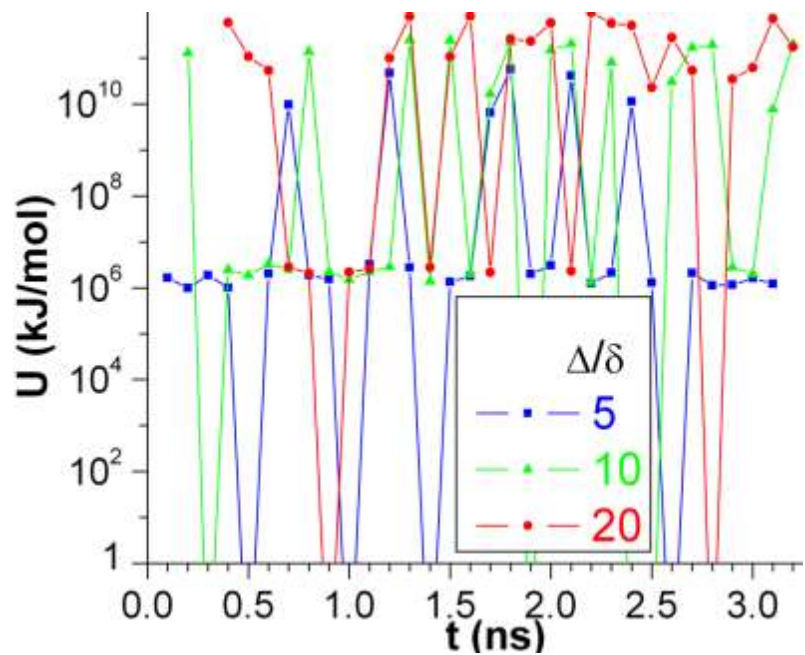

Fig. S4. Potential energy of EM configurations at different stages of 1PRT evolution. Large negative $U$ in some time points is an artefact and requires different set of Gromacs EM parameters.

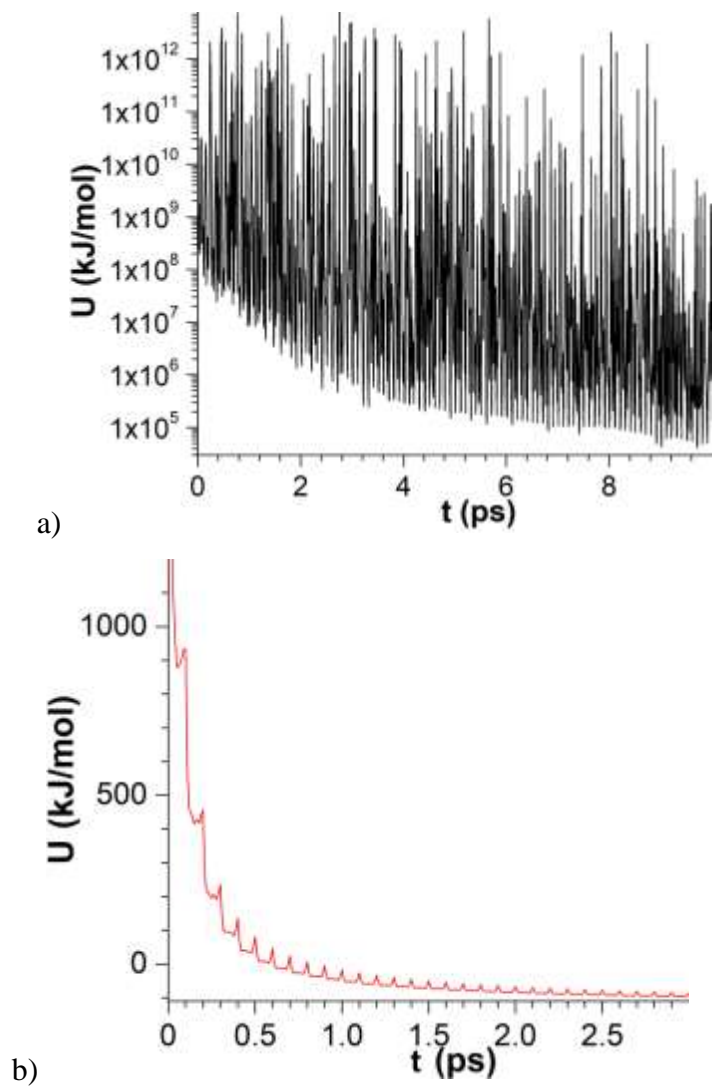

Fig. S5. Potential energy of 1PRT in energy minimization stage for $\Delta/\delta = 20$, extrapolated time $t = 400$ ps. a) For conventionally used Gromacs minimization parameters (Table I). Slow convergence of $U$ shows necessity to adjust the MD parameters to achieve a robust virtual MD protocol for this system. b) With cutoff distances for the short-range neighbor list and Coulomb and VdW radii increased from 1.4 to 2.0 nm.

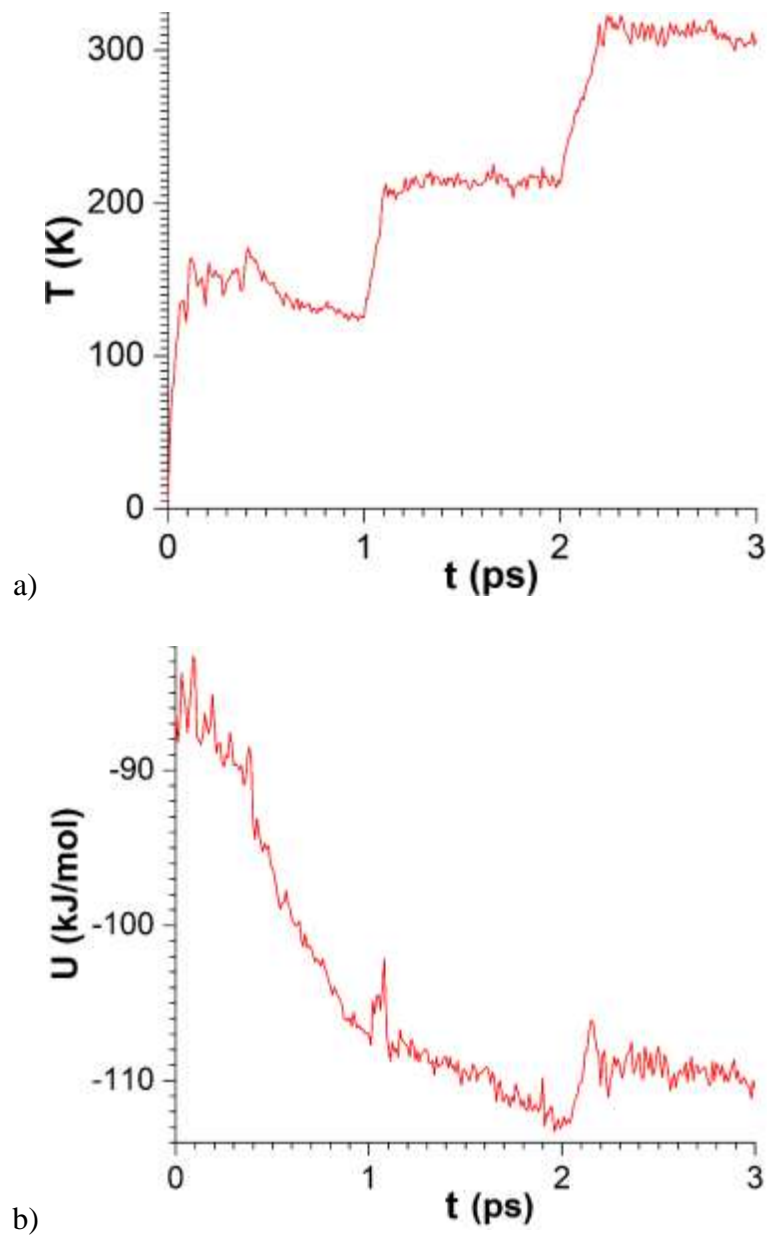

Fig. S6. During thermal equilibration of the NVE configuration, the temperature was increased by 100 K every 1 ps to arrive at a stable EQ configuration. $\Delta/\delta = 20$, $t = 400$ ps. a) Temperature, and b) potential energy.

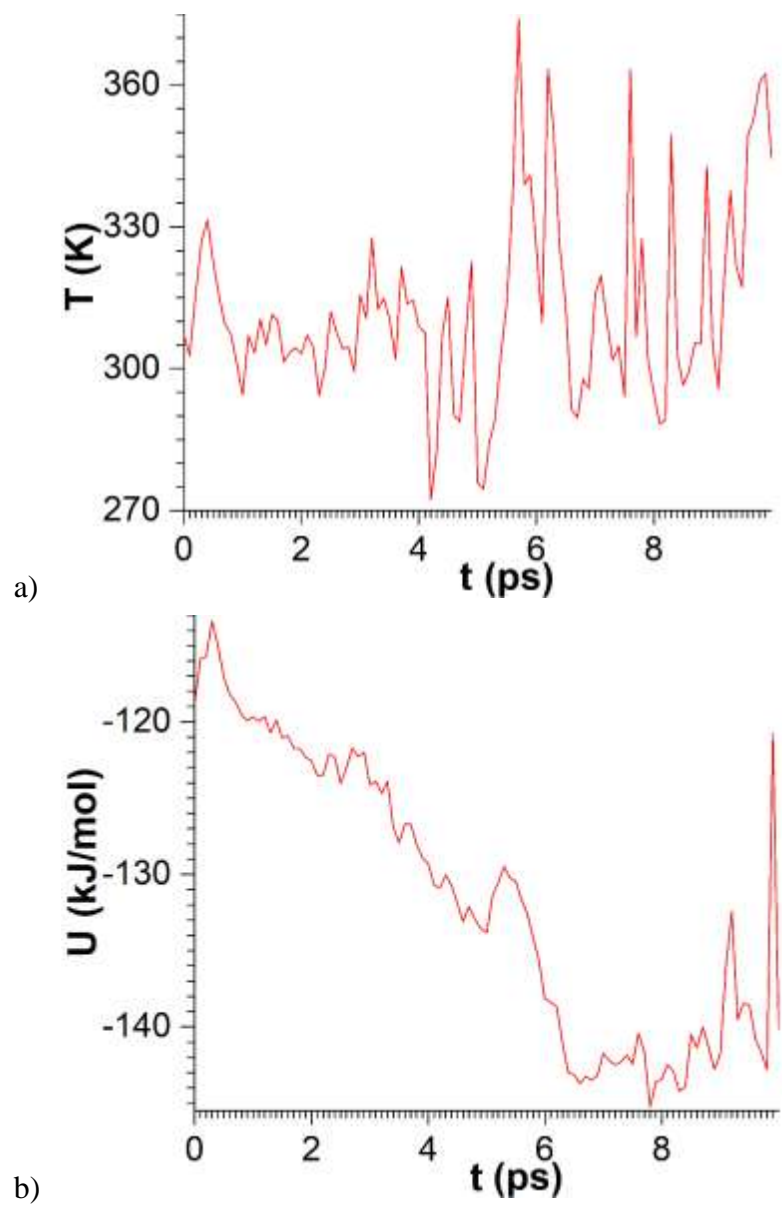

Fig. S7. Validation MD NVT run (first stage of the 2$^{nd}$ virtual MD step) for the final EQ configuration of Fig. S6. a) Temperature, and b) potential energy.

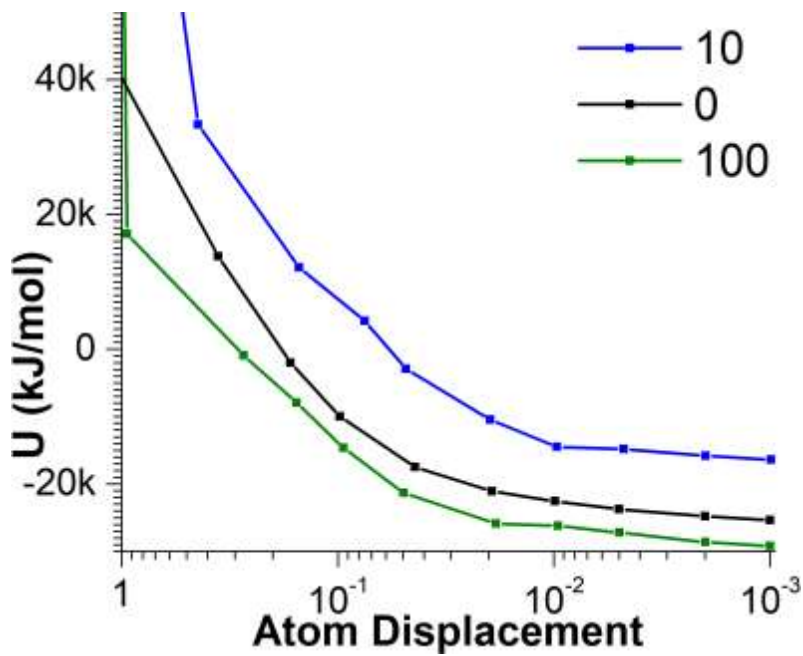

Fig. S8. Potential energy of the 1PRT protein PA configuration subjected to different amounts of MSR-assisted energy minimization as a function of the average atom displacement over the last MSR iteration for different starting points in conventional MD (shown in ps the *legend*). Achieving smaller displacements requires more MSR iterations, so that the potential energy becomes lower.

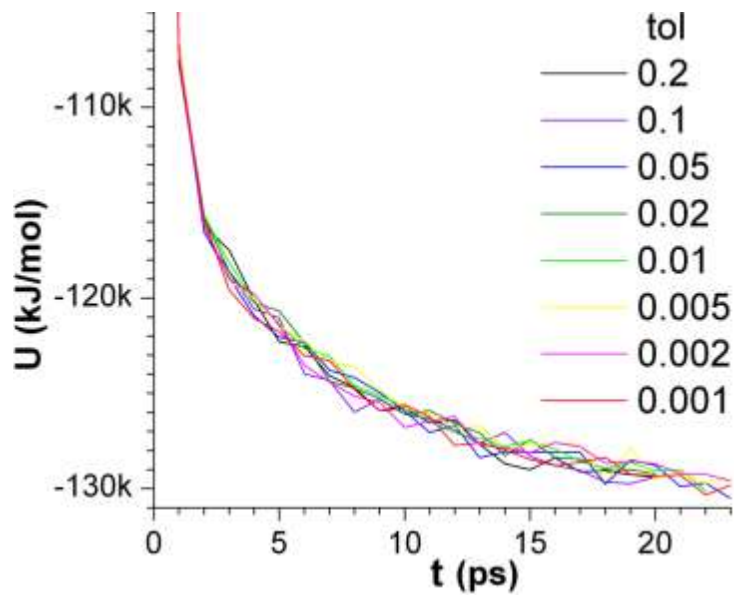

Fig. S9. Potential energy of 1PRT in course of its Gromacs MD simulation for different number of MSR iterations (shown in *legend* is the tolerance parameter which controls the number of iterations). Initial history time is $t - 2\Delta = 0$.

Apparently, the number of MSR iterations does not affect the course of MD simulation starting from the Pade-MSR configurations. This suggests necessity of optimization of the placement of MSR within the combined conventional – MSR minimization stage simultaneously with the duration of MSR.

**Coalescence of argon nanodroplets**

System of two identical spherical argon droplets with the radius R = 7.5 nm and the initial distance of 1 nm between droplets (Fig. 3) was prepared in VMD using equilibrated cubic box with 15 nm side length and the density of 1650 g/L. The energy minimization of the cube was skipped since it has converged in one step. Its equilibration was performed via MD NVT for 10 ps, during which the temperature was risen from 0 to 70 K and equilibrated. The potential energy was growing during heating up stage for ~1.8 ps, followed by monotonous decrease below initial value (data not shown). In the beginning of conventional MD NVT simulation, the potential energy fluctuates due to the formation of bridge between nanodroplets. After ~ 100 ps, the fluctuations become small and the energy slowly decreases (Fig. S10).

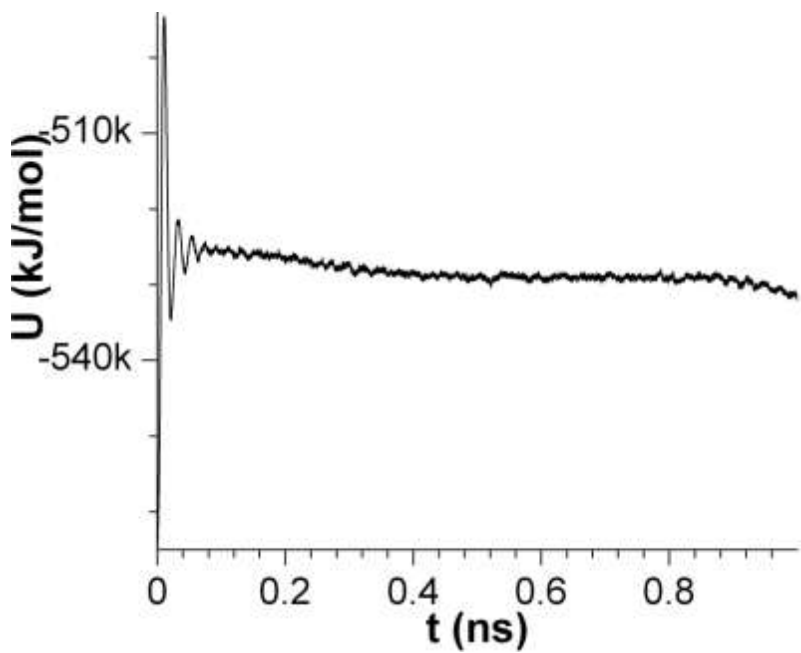

Fig. S10. Potential energy of the two argon nanodroplets coalescing during conventional MD NVT simulation in vacuum. Faster decrease of $U$ after 0.8 ns is due to attraction of the merged droplet to its two neighboring mirror images due to periodic boundary effects.

PA structures expectedly have high potential energy ($U$) due to the non-bonded, and thus more chaotic, liquid-like character of atom motions (Fig. S11).

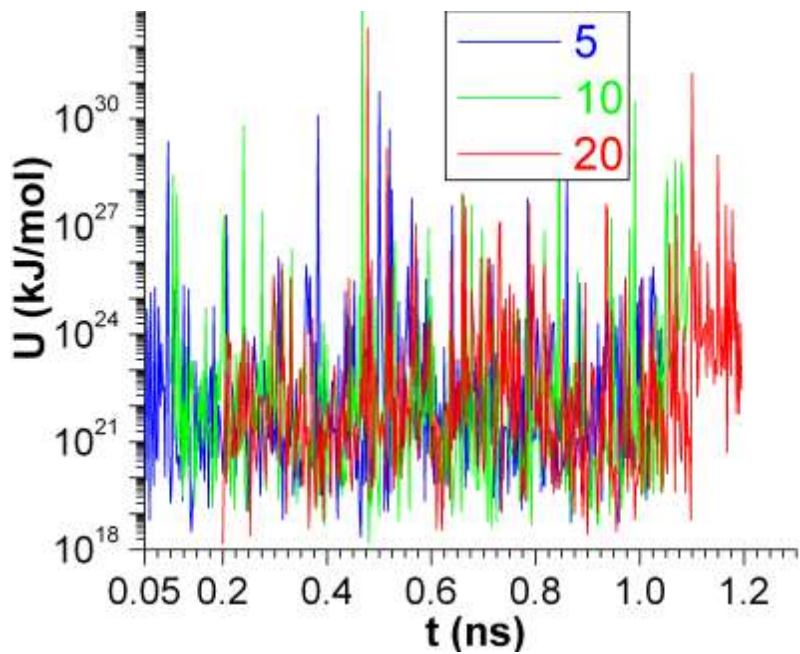

Fig. S11. Potential energy of the PA configurations for the two coalescing argon nanodroplets. *Legend* shows $\Delta/\delta$ values. Stride in time is 2.5 ps.

Conventional energy minimization of the PA configurations is illustrated in Fig. S12 and S13. Larger $\Delta/\delta$ values in the PA stage lead to higher $U$ (compare *solid lines* in Fig. S12) after sufficiently long energy minimization (for 1000 steps).

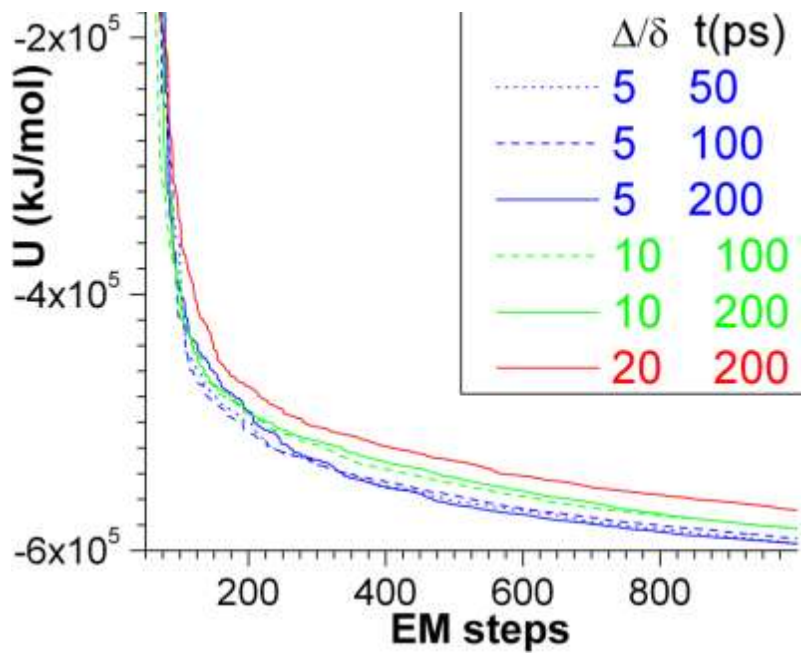

Fig. S12. Energy minimization of the PA configurations for two coalescing argon nanodroplets. *Colors* discriminate different $\Delta/\delta$ values in the PA, and *line styles* discriminate values of the system evolution time, as in the *legend*.

$U(t)$ after 1000 EM steps increases with the increase of $\Delta/\delta$ (Fig. S13).

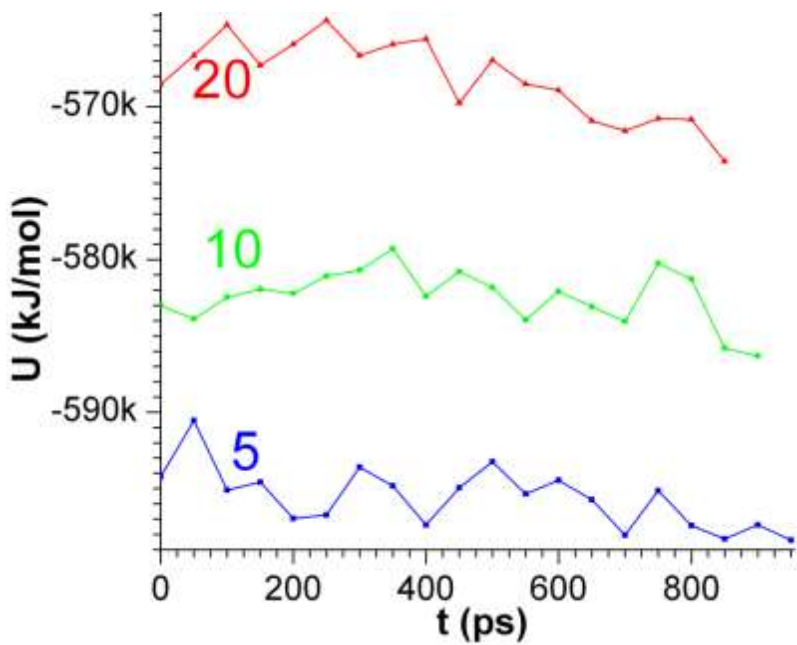

Fig. S13. Potential energy after 1000 steps of conventional energy minimization at different stages of argon system evolution. Stride in time is 50 ps. Notice the increase in $U$ at increasing $\Delta/\delta$, *labeled* on the plot.

Thermal equilibration using MD NVT makes $U$ converging close to its values in conventional MD (Fig. S14). Also, the equilibrated $U$ becomes lower in course of system evolution, as expected, thereby validating the virtual MD workflow.

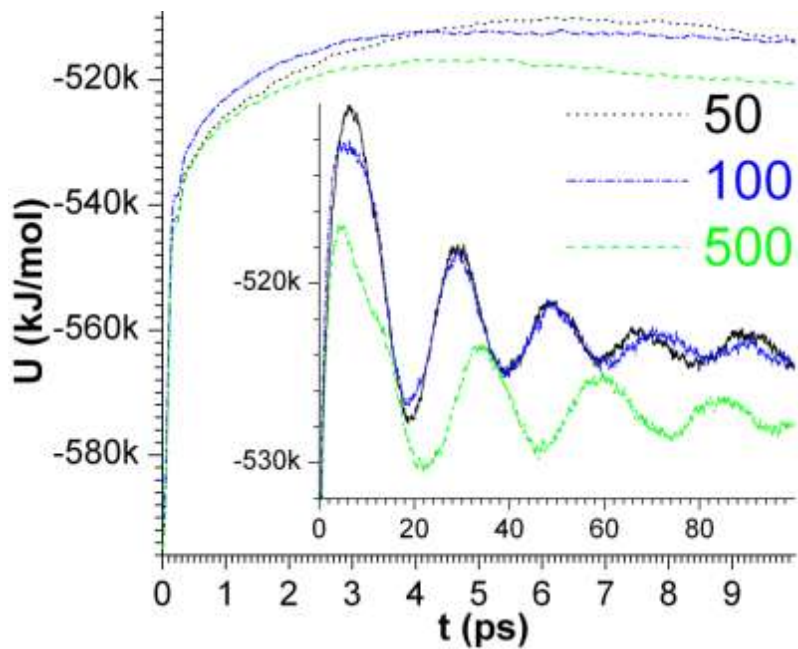

a)

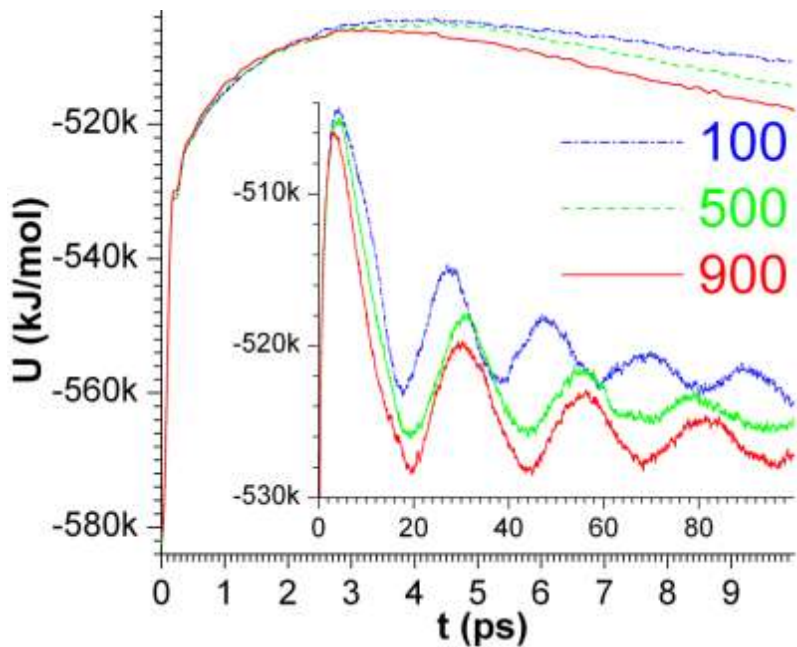

b)

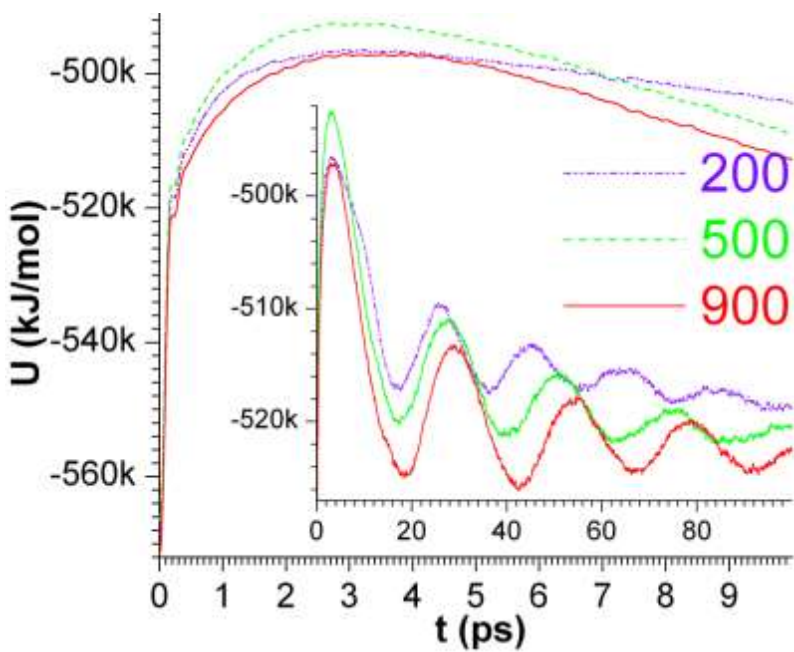

c)

Fig. S14. Potential energy course during MD NVT simulation stage of the second virtual MD step, extended to 100 ps (*insert*). Proximity of the converging $U$ to its value in conventional MD validates the first virtual MD step. The NVT simulation was launched after 1000 energy minimization steps starting at the PA configurations for a) $\Delta/\delta = 5$, b) $\Delta/\delta = 10$, c) $\Delta/\delta = 20$. PA extrapolated time $t$ is shown in the *legends*.

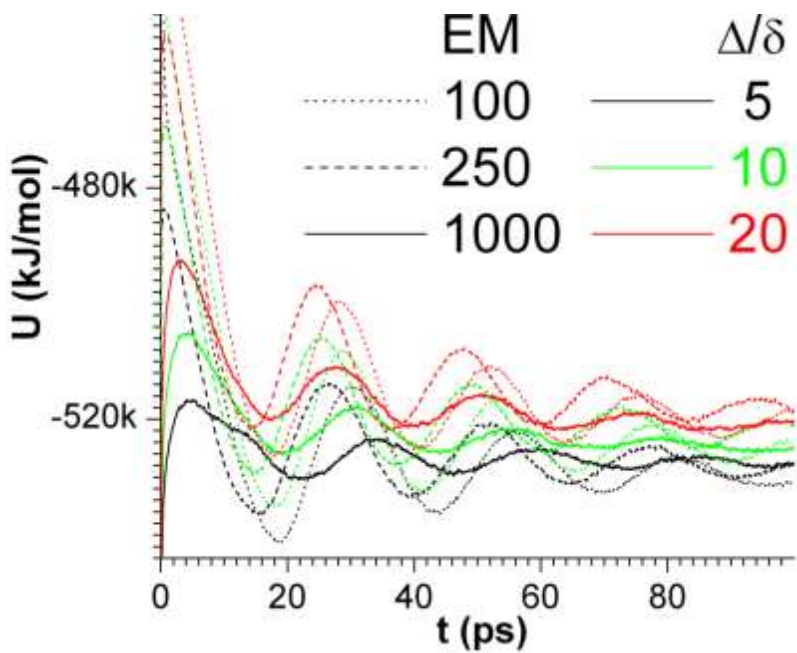

Fig. S15. Potential energy course during MD NVT simulation stage of the second virtual MD step, extended to 100 ps, for different number of energy minimization steps (indicated by *line styles*, see *legend*) and $\Delta/\delta$ (*colors*). All curves converge to the same energy level for a given ratio $\Delta/\delta$, and the levels are close to each other for different $\Delta/\delta$, proving robustness of the virtual MD algorithm with respect to the duration of minimization stage and the ratio $\Delta/\delta$.

Table S2 shows the duration of NVT equilibration simulation needed to reach the energy level of conventional MD for different $\Delta/\delta$ at different stages of system evolution. Clearly, more than 250 minimization steps should be taken to make energy equilibration short enough to achieve speedup in virtual MD. Thus, 100 minimization steps would require at least 11 ps of NVT simulation to reach the same energy level as in conventional MD, and 9.5 ps for 250 minimization steps. Deviations from 70 K upon reaching MD energy level do not exceed 0.5 K for 100 minimization steps and 0.3 K for 250 minimization steps, which means that the temperature has equilibrated. For 1000 minimization steps, the time to reach energy level of MD strongly decreases, but additional NVT equilibration is required to stabilize the temperature. Overall, the minimum duration of NVT is determined by energy and temperature equilibration, whichever is longer.

Table S2. Amount of time required for energy convergence and thermal equilibration of EM configuration during MD NVT simulation (starting with velocities generated for 70 K, as in the 1$^{st}$ stage of the 2$^{nd}$ virtual MD step). Time instances ($t$, in ps) and temperatures (T, in K) are shown when $U$ in the NVT becomes equal to that in conventional MD at time $t$, or approaches closest to it. The values of $U$ in conventional MD at corresponding times are -524741 kJ/mol at 50 ps, -525607 kJ/mol at 100 ps, -526981 kJ/mol at 200 ps, -528974 kJ/mol at 500 ps, and -529211 kJ/mol at 900 ps. Last row of the table shows the approximate time of thermal equilibration for the EM structure obtained with 1000 minimization steps. It is chosen to be the time when temperature deviation from the target value becomes less than 0.3 K for the first time after experiencing large fluctuations.

| EM | | $\Delta/\delta=5$ | | | $\Delta/\delta=10$ | | | $\Delta/\delta=20$ | | |
|---|---|---|---|---|---|---|---|---|---|---|
| | $t$ | 50 | 100 | 500 | 100 | 500 | 900 | 200 | 500 | 900 |
| 100 | $t$ | 10.965 | 11.422 | 12.894 | 12.702 | 14.249 | 14.241 | 16.090 | 17.726 | 16.132 |
| | T | 70.148 | 70.357 | 70.299 | 70.506 | 70.131 | 70.157 | 69.904 | 70.314 | 70.088 |
| 250 | $t$ | 9.479 | 10.191 | 10.448 | 12.195 | 13.225 | 12.575 | 13.135 | 14.125 | 15.474 |
| | T | 70.308 | 70.099 | 70.307 | 70.092 | 70.197 | 70.015 | 70.132 | 70.171 | 69.981 |
| 1000 | $t$ | 1.134 | 0.737 | 0.7745 | 0.334 | 0.284 | 0.265 | 0.132 | 0.1205 | 0.1345 |
| | T | 69.018 | 68.621 | 68.810 | 61.837 | 60.832 | 60.015 | 46.174 | 47.173 | 46.036 |
| | $t$ eq. | 1.6 | 1.6 | 1.7 | 1.9 | 1.6 | 1.6 | 1.6 | 1.6 | 1.7 |

Analysis of RMSD from the conventional MD simulation was performed for the microstates obtained during virtual MD stages. A symmetric RMSD using particle indistinguishability, as above (Sect. III), is considered below.

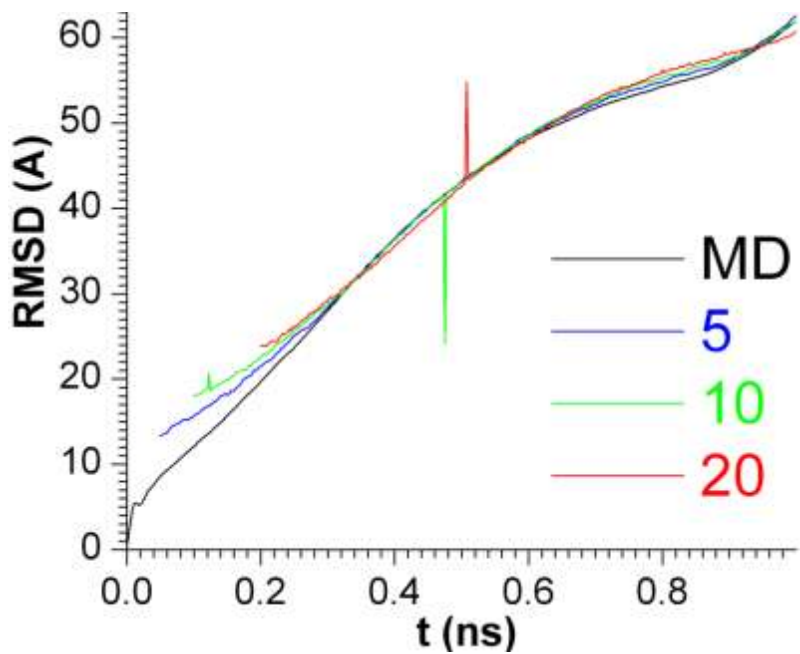

Fig. S16. RMSD of MD (*black*) and PA (*color*) configurations from initial state in conventional MD versus evolution time. $\Delta/\delta$ is distinguished by *colors*, see *legend*.

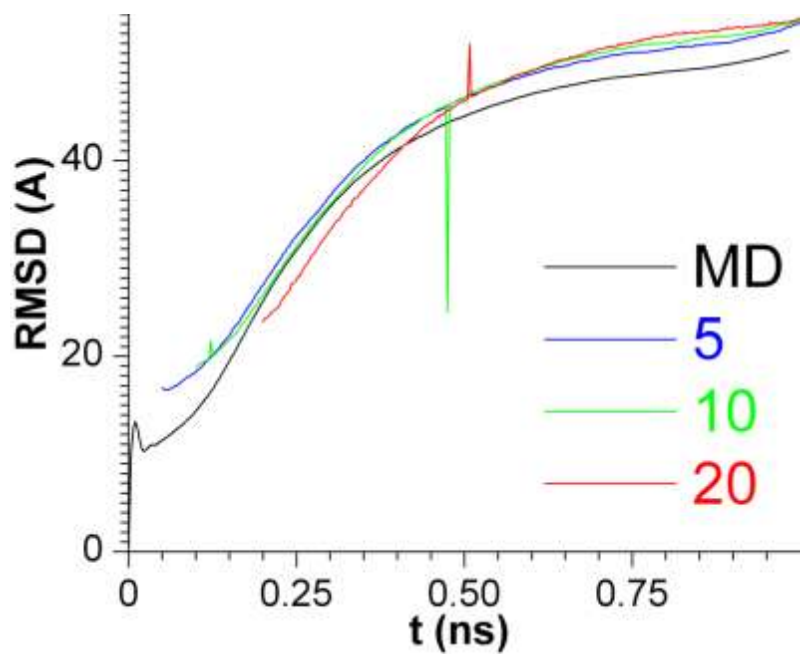

Fig. S17. RMSD accounting for particle indistinguishability for conventional MD and PA configurations from initial MD state. $\Delta/\delta$ is distinguished by *colors*, see *legend*.

Inclusion of NVT equilibration stage does not considerably affect the RMSD relative to conventional MD. In most cases, 1.6 ps of equilibration lowers the RMSD, and 11 ps makes it higher than that for the EM configuration (Fig. S18). At later stages of argon droplet coalescence the quality of microstate obtained after one virtual MD step improves (Fig. S18). Overall, the RMSD analysis might be used to refine the values of virtual MD parameters, such as the duration of thermal equilibration.

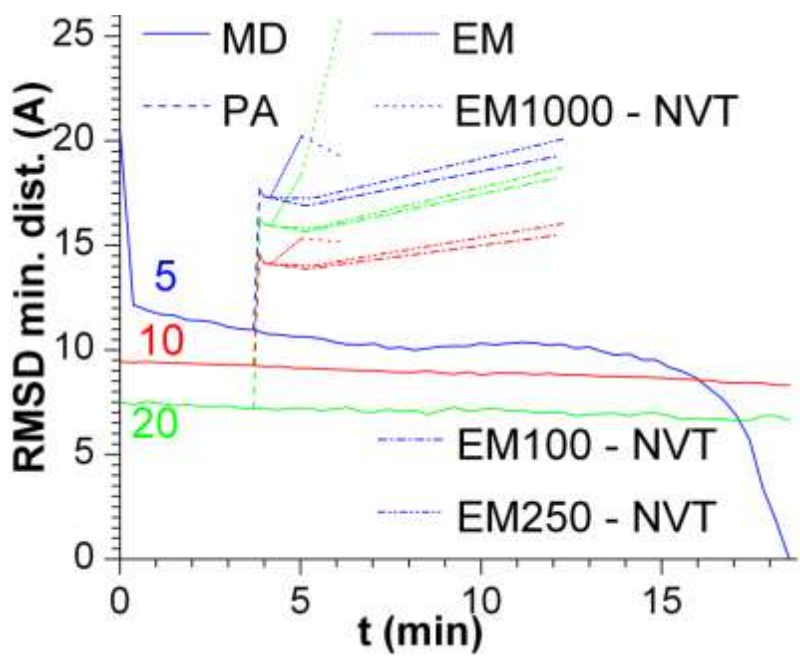

Fig. S18. RMSD with account for the indistinguishability of argon atoms at each stage of the virtual MD step for different ratios $\Delta/\delta$ at initial history time $t-2\Delta=0$. Extrapolated time $t=50\,ps$ for $\Delta/\delta=5$, $t=100\,ps$ for $\Delta/\delta=10$, and $t=200\,ps$ for $\Delta/\delta=20$. Ratios $\Delta/\delta$ are distinguished by *colors* and are *labeled* on the plot. Virtual MD stages are distinguished by *line styles*, see *legend* (e.g., "EM" means conventional energy minimization, "EM1000 - NVT" stands for NVT after 1000 minimization steps). Horizontal axis shows actual compute time in minutes. Shortest possible compute time is taken for each stage; it is achieved when using one compute node with 16 CPUs. RMSD for conventional MD simulation becomes zero after reaching duration $t=\Delta$.

The following actual compute times for virtual MD stages are involved in the Fig. S18. 5 ps of MD NVT takes 222.5 s, PA stage is estimated to take 9.2 s for the studied argon droplets system, 100 minimization steps take 7 s, 1.6 ps of thermal equilibration takes 71.2 s. Below the benchmarks are presented for the EM and MD NVT stages using an IBM NeXtScale nx360 M4 server equipped with two Intel Xeon E5-2650 v2 8-core processors. Number of CPUS for calculation of the speedup of virtual MD was taken based on minimal times for respective stages.

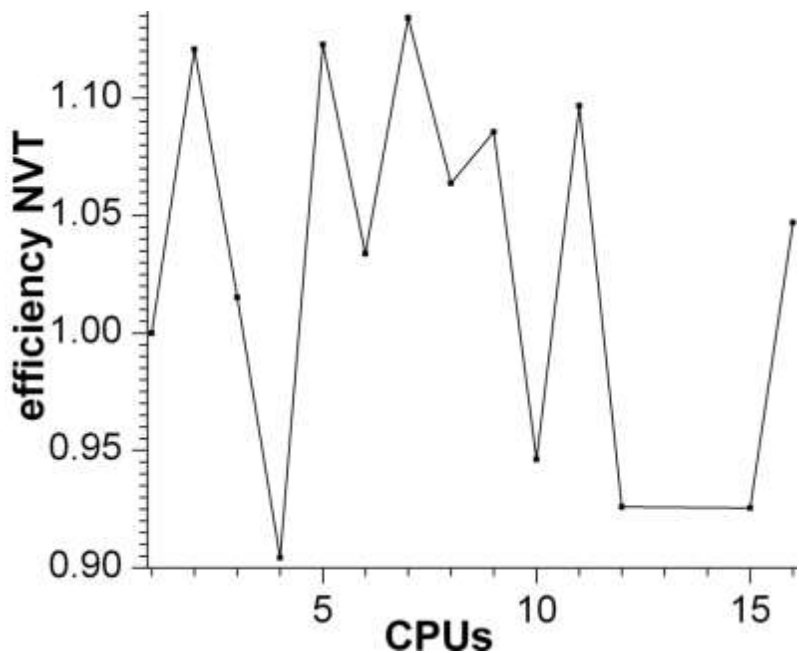

Fig. S19. Efficiency of MD NVT stage of a virtual MD step for upto 16 CPUs on one compute node relative to one CPU. The shortest simulation time is achieved for 16 CPUs, with the efficiency 1.047. Performance measurement unit is 4.829 ps/hr/CPU.

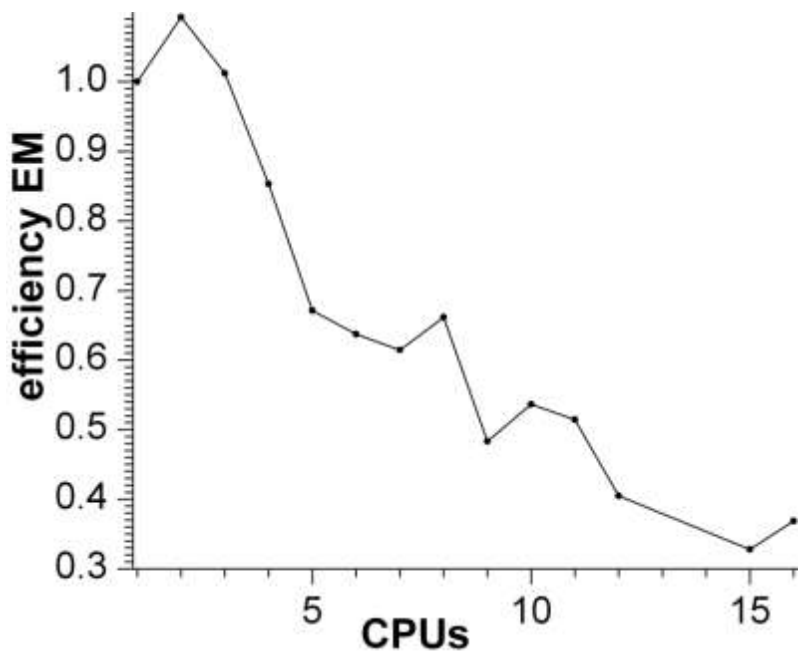

Fig. S20. Efficiency of EM stage of a virtual MD step for upto 16 cores on one compute node relative to one core. Performance measurement unit is 2.421 EM steps/s/CPU. Highest efficiency of EM is achieved for 2 CPUs, the shortest simulation time – for 16 CPUs.